\documentclass[12pt]{article}

\usepackage[a4paper,left=30mm,right=20mm,top=20mm,bottom=20mm]{geometry}
\usepackage{graphics}
\usepackage{amsmath}
\usepackage{epsfig}
\usepackage{cite}

\title{Non-prompt $J/\psi + J/\psi$ and $J/\psi + \mu$ production at LHC as a test for TMD gluon density in a proton}
\title{Associated non-prompt $J/\psi + \mu$ and $J/\psi + J/\psi$ production at LHC as a test for TMD gluon density}
\author{S.P.~Baranov$^{1}$, A.V.~Lipatov$^{2,3}$, M.A.~Malyshev$^{2}$}

\begin{document}

\maketitle

\begin{center}
{\it $^1$P.N.~Lebedev Physics Institute, 119991 Moscow, Russia}\\
{\it $^2$Skobeltsyn Institute of Nuclear Physics, Moscow State University, 119991 Moscow, Russia}\\
{\it $^3$Joint Institute for Nuclear Research, Dubna 141980, Moscow region, Russia}
\end{center}

\vspace{0.5cm}

\begin{center}

{\bf Abstract }

\end{center} 

\indent

We consider the associated production of $J/\psi$ mesons and muons
originating from the $b$-flavored hadron decays and non-prompt double $J/\psi$ production at the LHC
using the $k_T$-factorization approach.
To describe the inclusive 
$b$-hadron decays into the different charmonium states
we apply fragmentation approach 
and adopt fragmentation functions based on the 
non-relativistic QCD factorization.
The transverse momentum dependent (TMD) gluon densities in a proton are determined 
using the Catani-Ciafaloni-Fiorani-Marchesini equation and
Kimber-Martin-Ryskin prescription. 
We investigate the effects coming from parton showers,
estimate the double parton scattering contribution
and compare our predictions with the first experimental
data taken by the ATLAS and LHCb Collaborations at $\sqrt s = 8$~TeV.
These data can serve as an additional test for TMD gluon density function in a proton.

\vspace{1.0cm}

\noindent
PACS number(s): 12.38.-t, 12.38.Bx, 13.25.Hw, 14.40.Pq

\newpage

Very recently, the ATLAS Collaboration has presented a new measurement of 
the production of two $b$-hadrons at $\sqrt s = 8$~TeV, where one of these $b$-hadrons decays into a $J/\psi$ meson (with 
the subsequent decay of the latter into muon pair) and the other decays into $\mu + X$, 
resulting in three muons in the final state\cite{1}.
The kinematic correlations for pairs of $b$-hadrons,
reconstructed via their inclusive decays 
into $J/\psi$ mesons, have been also measured by the LHCb Collaboration in the 
forward rapidity region $2 < y^{J/\psi} < 4.5$\cite{2}.
These data provide an additional testing ground for perturbative
Quantum Chromodynamics (pQCD) predictions of heavy flavour 
production, especially at small opening angles between the heavy hadrons, where the relevant
theoretical uncertainties are rather large.
A number of differential cross sections, 
including different angular correlations between the final decay muons (or rather $J/\psi$ mesons)
was measured for the first time, that stimulated us to 
perform corresponding calculations in the framework of the $k_T$-factorization
approach\cite{3,4} and compare these predictions with the ATLAS and LHCb data\cite{1,2}.
The $k_T$-factorization approach is mainly based on the 
Balitsky-Fadin-Kuraev-Lipatov\cite{5} (BFKL) 
or Catani-Ciafaloni-Fiorani-Marchesini\cite{6} (CCFM)
gluon dynamics at small $x$ and has certain technical advantages in the 
ease of including higher-order radiative corrections that can be taken into account 
in the form of transverse momentum dependent (TMD) 
parton distributions\footnote{See reviews\cite{7,8} for more 
information.}. It has become a widely 
exploited tool and it is of interest and importance to test it in as many cases as 
possible. Closely related to this is selection of the TMD parton densities best 
suited to describe data\footnote{It was shown\cite{9} that one can 
reconstruct the full map of the TMD parton 
densities in a proton by applying
different cuts on the final $b \bar b$ states.}. These tasks form the major goal of our present article. 
Additionally, we investigate the influence of parton showers on description of the ATLAS and LHCb data\cite{1,2}
and estimate the contributions from the double parton scattering (DPS) mechanism,
now widely discussed in the literature.
The consideration below continues the line of our previous studies\cite{9,10}, where we 
inspect the evolution details of $b$-quarks fragmenting and decaying into final state charmonia.

The calculations of non-prompt $J/\psi$ meson and a muon associated production
(and non-prompt $J/\psi + J/\psi$ production)
involve several main ingredients: the cross sections of $b$-hadron production,
the partial widths of their subsequent decays into the different charmonia states 
and/or semileptonic decays of $b$-hadrons.
Below we collect the previously tested
components of the theory and only briefly recall our main points.

First, to calculate the cross sections of inclusive $b$-hadron production 
in $pp$ collisions we apply
the $k_T$-factorization approach,
mainly based on the ${\cal O}(\alpha_s^2)$ off-shell gluon-gluon fusion subprocess:
\begin{equation}
  g^*(k_1) + g^*(k_2) \to b(p_1) + \bar b (p_2),
\end{equation}

\noindent
where the four-momenta of all particles are given in
the parentheses.
The corresponding gauge-invariant off-shell (dependent on transverse momenta 
of the initial gluons) production amplitude was calculated earlier (see, for example,\cite{11} and references therein).
Then, $b$-flavor production cross section can be obtained as a convolution of the off-shell 
partonic cross section $\hat \sigma_{gg}^*(x_1,x_2,{\mathbf {k}_{1T}^2},{\mathbf {k}_{2T}^2},\mu^2)$ and the TMD gluon 
distributions in a proton $f_g(x,{\mathbf {k}_{T}^2},\mu^2)$:
\begin{equation}
  \sigma = \int dx_1 dx_2 \, d{\mathbf {k}_{1T}^2} d{\mathbf {k}_{2T}^2} \, d\hat \sigma_{gg}^*(x_1,x_2,{\mathbf {k}_{1T}^2},{\mathbf {k}_{2T}^2},\mu^2) f_g(x_1,{\mathbf {k}_{1T}^2},\mu^2)  f_g(x_2,{\mathbf {k}_{2T}^2},\mu^2),
\end{equation}

\noindent
where ${\mathbf k}_{i\,T}$ being the component of the off-shell 
gluon momentum $k_i$ perpendicular
to the beam axis ($k_i^2 = - {\mathbf k}_{iT}^2 \neq 0$, $i = 1$ or~$2$), $x_i$ is the fraction of 
longitudinal momentum of the colliding proton and $\mu^2$ is the hard scale.
The subsequent fragmentation of the produced $b$ quarks into $b$-hadrons is described with the
Peterson fragmentation function\cite{12} with $\epsilon_b = 0.0126$.
The consistency of this setting was shown in a previous paper\cite{10}.
We have tested two families of 
the TMD gluon distribution functions in a proton, which are widely discussed in the literature and
often used in appications. 
So, we used a numerical solution\cite{13}
of the CCFM equation (labeled below as {\it JH'2013} family).
The CCFM gluon evolution equation 
provides a suitable tool since it smoothly interpolates 
between the small-$x$ BFKL gluon dynamics and high-$x$ DGLAP dynamics.
Two sets of the TMD gluon densities were determined from the fits 
to high precision HERA data on the proton structure functions: {\it JH'2013 set 1}, which was 
determined from a fit to inclusive $F_2(x, Q^2)$ data only, and
{\it JH'2013 set 2}, which was determined from a fit to both $F_2(x, Q^2)$
and $F_2^c(x, Q^2)$.
As an alternative choice, we applied the TMD gluon density obtained
from the Kimber-Martin-Ryskin\cite{14,15} (KMR) prescription.
The KMR approach is a formalism to construct the TMD parton (quark and gluon)
densities from well-known conventional ones,
developed at leading order\cite{14} (LO) and next-to-leading order\cite{15} (NLO). 
The key assumption of this approach is that the
$k_T$-dependence of the TMD parton distributions enters at the last
evolution step, so that the usual DGLAP evolution can be used up to this step.
For the input, we used Martin-Stirling-Thorn-Watt set\cite{16} ({\it MSTW'2008}) at LO and NLO, respectively. 
The phenomenological consequences of our different choices for the TMD gluon densities in a proton
are discussed below\footnote{At the moment, there is a large variety of the TMD gluon distribution functions
in a proton available. Most of them are collected in the \textsc{tmdlib} package\cite{17}, which is a C++ library 
providing a framework and an interface to the different parametrizations.}.

Next step of our calculations is connected with the description of the $J/\psi$ production from $b$-hadron decays
("non-prompt" production). 
We adopted the so-called fragmentation approach, as it was 
done earlier\cite{10}.
In this approach, the calculated $b$-hadron cross section has to be 
convoluted with $b\to J/\psi + X$ fragmentation function, which is 
the longitudinal momentum distribution of the $J/\psi$ meson 
from $b$-hadron decay, appropriately boosted along 
the $b$-hadron flight direction.
The latter have been obtained\cite{18} in the framework of the 
nonrelativistic QCD (NRQCD) factorization\cite{19,20} using the approach\cite{21}
and reasonably agree with the
CLEO\cite{22} and BABAR\cite{23} measurements.
The formalism\cite{18} was implemented into our calculations
without any changes (see\cite{10} for more details).
To be precise, we employ the asymptotic expression\cite{18} for the $b$-hadron decay
distribution differential in the longitudinal momentum fraction $z$
carried by the produced charmonium state, obtained in the limit $|{\mathbf p}_b| \gg m_b$,
where $p_b$ and $m_b$ are the momentum and the mass of decaying $b$-hadron.
This approximation is valid within $11$\% and $5$\% accuracy 
for $|{\mathbf p}_b| = 10$~GeV and $20$~GeV, respectively, that is
suitable for our phenomenological study.
According to the ATLAS and LHCb experimental setup\cite{1,2}, we also took into account
feed-down contributions from the excited charmonium states, namely,
$b \to \chi_{cJ} + X$ (with $J = 0, 1, 2$) and $b \to \psi(2S) + X$ decays
followed by subsequent radiative decays $\chi_{cJ} \to J/\psi + \gamma$ 
and $\psi(2S) \to J/\psi + \gamma$ using the same 
approach\cite{18}.
We set the branching fractions $B(b \to J/\psi + X) = 0.68$\%,
$B(b \to \psi(2S) + X) = 0.18$\%, $B(b \to \chi_{c0} + X) = 0.015$\%,
$B(b \to \chi_{c1} + X) = 0.21$\%, $B(b \to \chi_{c2} + X) = 0.026$\%, 
$B(\psi(2S) \to J/\psi + \gamma) = 61$\%,
$B(\chi_{c0} \to J/\psi + \gamma) = 1.27$\%,
$B(\chi_{c1} \to J/\psi + \gamma) = 33.9$\%,
$B(\chi_{c2} \to J/\psi + \gamma) = 19.2$\%\cite{24}.
Finally, $J/\psi \to \mu^+ \mu^-$ decays were generated according to the phase space
with the branching fraction $B(J/\psi \to \mu^+ \mu^-) = 5.961$\%\cite{24}.

To produce muons from $b$-hadron decays, 
we simulate their semileptonic decay according to the standard 
electroweak theory with the branching fraction $B(b \to \mu) = 9.8$\%\cite{24}. We took into account also muons produced 
in semileptonic cascade decays (the decay of a $c$-hadron produced 
in the decay of a $b$-hadron) with $B(b\to c\to \mu) = 8.02$\%\cite{24}.
Other essential parameters, such as renormalization and factorization scales, masses of produced particles are
taken exactly the same as in our previous studies\cite{10}. 

We close the short description of our calculation steps with DPS contributions,
where we apply a simple factorization formula (for details see the reviews\cite{25,26,27} 
and references
therein):
\begin{equation}
  \sigma_{\rm DPS}(J/\psi + \mu) = {\sigma(J/\psi)\,\sigma(\mu) \over \sigma_{\rm eff}},
\end{equation}

\noindent 
where $\sigma_{\rm eff}$ is a normalization constant which incorporates 
all ``DPS unknowns'' into a single phenomenological parameter.
A similar expression (with an extra factor of $1/2$ due to identity of final state particles)
is valid in the case of non-prompt $J/\psi + J/\psi$ production.
A numerical value of $\sigma_{\rm eff} \simeq 15$~mb has been obtained
from fits to $pp$ and $p\bar p$ data (see, for example,\cite{28})
and will be taken as the default value throughout the paper.
Note that one can easily estimate the relative DPS contribution to the
total cross section as $\sigma_{\rm DPS}/\sigma_{\rm SPS} \sim \sigma(b\bar b)/\sigma_{\rm eff}$.
This value could be non-zero and DPS could contribute 
significantly in some kinematical regions.
Below we will clarify this point for both considered processes.

We discuss first the associated non-prompt 
$J/\psi + \mu$ production at $\sqrt s = 8$~TeV.
The ATLAS Collaboration has measured the corresponding total and differential cross sections
in a restricted part of the phase 
space (fiducial volume). So, each muon was required to
have transverse momentum $p_T > 6$~GeV, the two muons originating from the 
$J/\psi$ decay must have pseudorapidities $|\eta| < 2.3$ and the third muon 
must have $|\eta| < 2.5$\cite{1}.  We implemented the experimental setup used by the ATLAS Collaboration 
in our numerical program. 
Several normalized differential cross sections have been presented for the first time, namely: 
transverse momentum of the three-muon system $p_T(J/\psi,\mu)$, 
azimuthal separation between the $J/\psi$ and third muon $\Delta \phi(J/\psi, \mu)$,
separation between the $J/\psi$ and third muon in the azimuth-rapidity plane $\Delta R(J/\psi, \mu)$,
separation in rapidity between the $J/\psi$ and third muon $\Delta y(J/\psi, \mu)$,
magnitude of the average rapidity of the $J/\psi$ and third muon $y_{\rm boost}(J/\psi, \mu)$,
mass of the three-muon system $M(J/\psi,\mu)$, 
ratio of the invariant mass of three-muon system to the transverse momentum of three-muon system $m^{\mu\mu\mu}/p_T^{\mu\mu\mu}$
and its inverse $p_T^{\mu\mu\mu}/m^{\mu\mu\mu}$.
In some sense, studying of the normalized differential cross sections could lead to a bit
more stringent comparison between data and theory due to reduced experimental (mainly systematic)
uncertainties.

We confront our predictions with the available data in Figs.~1 --- 3.
The solid histograms represent our central predictions calculated with  
fixed renormalization $\mu_R$ and factorization $\mu_F$ scales at their default 
values (see\cite{10} for the detailed description of our input), 
while the shaded regions correspond to scale uncertainties of our predictions.
In the case of CCFM-evolved gluon densities ({\it JH'2013} family), to estimate the latter we used the 
{\it JH'2013 set 1(2)+} and {\it JH'2013 set 1(2)--} sets instead of default {\it JH'2013 set 1(2)} distributions.
These sets represent a variation of the renormalization
scale used in the off-shell production amplitude. 
The {\it JH'2013 set 1(2)+} set stands for a variation of $2\mu_R$, while 
set {\it JH'2013 set 1(2)--} refects $\mu_R/2$ (see\cite{13}).
To estimate the scale uncertainties of the KMR predictions,
we have varied both renormalization and factorization scales around their default values.
As one can see, the calculated cross sections (except $y_{\rm boost}(J/\psi, \mu)$ spectrum)
strongly depend on the TMD gluon density used.
A clear difference in shape between the {\it JH'2013} and KMR predictions
is observed for angular correlations $\Delta R(J/\psi, \mu)$ and $\Delta \phi(J/\psi, \mu)$ (see Fig.~1)
and distributions on $p_T(J/\psi, \mu)$ and $M(J/\psi,\mu)$ (see Fig.~2).
A better description of all these observables is achieved with the KMR family of gluon 
distributions. There is only small overestimation of the data
at low $\Delta R(J/\psi, \mu)$ and $\Delta \phi(J/\psi, \mu)$ and in the 
last bins of $p_T(J/\psi, \mu)$ and $M(J/\psi,\mu)$, although the  
data are rather close to the estimated uncertainty bands.
In contrast, both the {\it JH'2013} gluon densities do not reproduce well
the measured shape of angular correlations: they underestimate 
the data at low $\Delta R(J/\psi, \mu)$ and $\Delta \phi(J/\psi, \mu)$ 
(especially at $p_T(J/\psi, \mu) > 20$~GeV) and tend to overestimate 
the data at $\Delta \phi(J/\psi, \mu) \sim \pi$ (see Fig.~1).
This is in agreement with the general trend observed in the $b\bar b$ di-jet production\cite{11}.
The measured $M(J/\psi,\mu)$ distribution is not reproduced with the CCFM-evolved
gluon densities.
While the $y_{\rm boost}(J/\psi, \mu)$ distribution is reasonably well described
within the theoretical and experimental uncertainties,
the $\Delta y(J/\psi, \mu)$ spectrum is somewhat poorly described by all gluon densities
under consideration with the tendency to fall away at high $\Delta y(J/\psi, \mu)$.
The predictions, obtained with the KMR gluon density, calculated with the LO and NLO accuracy,
are close to each other (and even coincide practically within the uncertainties).
The estimated DPS contributions are found to be small in the considered kinematic
region. Of course, some reasonable variations in $\sigma_{\rm eff} \simeq 15 \pm 5$~mb
would affect DPS predictions, though without changing our conclusion.

The observed difference between the KMR and {\it JH'2013} predictions for 
transverse momentum $p_T(J/\psi, \mu)$ and invariant mass $M(J/\psi,\mu)$ distributions 
leads to a noticable difference in the $m^{\mu\mu\mu}/p_T^{\mu\mu\mu}$ and its inverse $p_T^{\mu\mu\mu}/m^{\mu\mu\mu}$ 
spectra, as it is shown in Fig.~3.
Moreover, this difference becomes even more clearly pronounced: the ratio of the 
KMR and {\it JH'2013} predictions reaches $\sim 2.5 - 3$ at $p_T^{\mu\mu\mu}/m^{\mu\mu\mu} > 2$.
Therefore, such observables are particularly sensitive to the non-collinear gluon evolution dynamics
and, in addition to well-known properties of angular correlations between the momenta of the 
produced particles, could be very promising to constrain the TMD gluon densities in a proton.

Now we turn to the non-prompt $J/\psi + J/\psi$ production. The LHCb Collaboration
has presented\cite{2} the normalized cross sections measured as functions of several variables,
namely, the transverse momentum, rapidity and invariant mass of the $J/\psi$ pair,
the difference in the azimuthal angle between the momentum directions of two $J/\psi$ mesons,
the difference in the rapidity and pseudorapidity between them and 
the assymetry ${\cal A}_T$ between the transverse momenta of produced mesons:
\begin{equation}
  {\cal A} = \left| { p_T^{{J/\psi}_1} - p_T^{{J/\psi}_2} \over p_T^{{J/\psi}_1} + p_T^{{J/\psi}_2} } \right|.
\end{equation}

\noindent 
These data were obtained in the forward rapidity region $2 < y^{J/\psi} < 4.5$
for different requirements on the minimum transverse momentum of the $J/\psi$ mesons.
The results of our calculations are shown in Figs.~4 --- 10 in comparison with the LHCb data\cite{2}.
One can see that, in general, all the TMD gluon densities under consideration describe 
the data reasonably well for all distributions within the uncertainties.
However, both the CCFM-evolved gluons tend to overestimate the measured transverse momentum
spectra of the $J/\psi$ pair at low $p_T (J/\psi, J/\psi)$ and underestimate the invariant mass
distributions near the threshold, where $M(J/\psi, J/\psi) \leq 10$~GeV.
The KMR calculations agree well with the LHCb data for these observables.
We see again that a clear difference between the predictions 
is observed in the angular correlations (see Fig.~9).
Similar to the $J/\psi + \mu$ production,
both the {\it JH'2013} gluon densities do not reproduce the measured shape of 
$\Delta \phi(J/\psi,J/\psi)$ distributions: they underestimate 
the LHCb data at low $\Delta \phi(J/\psi,J/\psi)$ and overestimate 
the data at $\Delta \phi(J/\psi,J/\psi) \sim \pi$, although 
at large transverse momenta overall agreement becomes better.
In contrast, the both KMR gluons provide good description of these angular correlations.

The DPS contributions are also found to be small in the considered kinematic
region. Note that it was discussed earlier\cite{29} that a special correction
factor should be included into DPS calculations at LHCb conditions 
to take into account limited partonic phase space.
We have found that such a factor $F \sim (1 - x_1 - x_2)^2$
results in $\sim 15 - 20$\% smaller DPS cross sections (not shown in Figs.~1 --- 10),
that, in any case, does not change our conclusions.
For the ATLAS kinematical region the influence of this correction
factor is negligible.

As a last point of our study, we would like to note that the angular correlations in $b\bar b$ production
(and, therefore, $\Delta \phi(J/\psi, \mu)$, $\Delta R(J/\psi, \mu)$ and $\Delta \phi(J/\psi, J/\psi)$ 
distributions considered above)
are known to be sensitive to the inclusion of (initial and final state) parton
radiation that can either be simulated as parton showers or taken as an
additional higher-order process\footnote{A large piece of 
higher-order corrections containing leading $\log 1/x$ enhancement of cross sections due 
to real initial state parton emissions is already included in the our $k_T$-factorization 
calculations, as it was noted above.}. For the latter, we have estimated the partial contribution 
coming from the ${\cal O}(\alpha_s^3)$ quark-gluon scattering subprocess 
$q g \to q b\bar b$ within the conventional (collinear) QCD factorization 
using the \textsc{madgraph} tool\cite{30}. We found it to be
negligible everywhere (not shown in the Figs.~1 --- 10).
Additionally, we investigated the influence of parton showers
on the description of the ATLAS and LHCb data\cite{1,2}. For these studies we used 
a parton shower algorithm implemented in the Monte-Carlo event generator \textsc{cascade}\cite{31}.  
The results of our calculations for $\Delta \phi(J/\psi, \mu)$, $\Delta R(J/\psi, \mu)$ and 
$\Delta \phi(J/\psi, J/\psi)$ distributions
are shown in Figs.~11 and 12,
where the {\it JH'2013 set 2} gluon density is applied as an example.
We observed only a small effect at low $\Delta \phi(J/\psi, \mu)$ 
and $\Delta \phi(J/\psi, J/\psi)$
in the kinematical region of ATLAS and LHCb experiments\cite{1,2}.
This effect is coming from the final state parton showers only,
that could be 
easily understood since in our calculations the 
initial state parton emissions are already determined from 
the TMD gluon density and, therefore, 
do not influence the $k_T$ of the gluons.
However, parton showers could be very important at some kinematical regions (see, for example,\cite{32,33}).

To conclude, we applied the $k_T$-factorization approach
to investigate the 
associated production of $J/\psi$ mesons and muons
originating from the $b$-hadron decays 
and non-prompt $J/\psi + J/\psi$ production
at the LHC.
Our calculations were inspired by the recent experimental data presented by the 
ATLAS and LHCb Collaborations at $\sqrt s = 8$~TeV, where a number of 
differential cross sections were measured for the first time.
To describe the inclusive $b$-hadron decays into the $J/\psi$ mesons 
we used fragmentation approach with 
corresponding fragmentation functions calculated within the NRQCD.
Following the experimental setup, we 
took into account both direct $J/\psi$ production mechanism and feed-down contributions 
from the radiative decays of excited charmonium states.
The semileptonic $b$-hadron decays were 
generated according to the standard 
electroweak theory.
Numerically, we have tested two families of the TMD gluon densities in a proton,
namely, the CCFM-evolved gluon distributions ({\it JH'2013} sets) and the KMR ones,
calculated with the LO and NLO accuracy. 
Theoretical uncertainties and effects arising from parton showers, higher-order pQCD corrections 
and double parton scattering mechanism were estimated. 
Quite satisfactory agreement between the predictions and the data
can be regarded as another voice in support of the chosen TMD gluon
parametrizations. The tri-muon and $J/\psi + J/\psi$ measurements at the LHC did and will continue to play
their role in providing the useful and necessary experimental
constraints.

{\sl Acknowledgements.} 
We would like to thank H.~Jung for 
his extreme help in the calculation of the parton showers within the \textsc{cascade} Monte-Carlo event generator,
very useful discussions and important remarks.
We are very grateful to DESY Directorate for the 
support in the framework of Moscow --- DESY project on Monte-Carlo implementation for
HERA --- LHC. M.A.M. was also supported by a grant of the foundation for
the advancement of theoretical physics and mathematics "Basis" 17-14-455-1.
Part of this work was done by M.A.M. during his stay at DESY, 
funded by DAAD (Program "Research Stays for University Academics and Scientists").

\newpage 

\begin{figure}
\begin{center}
\includegraphics[width=7.6cm]{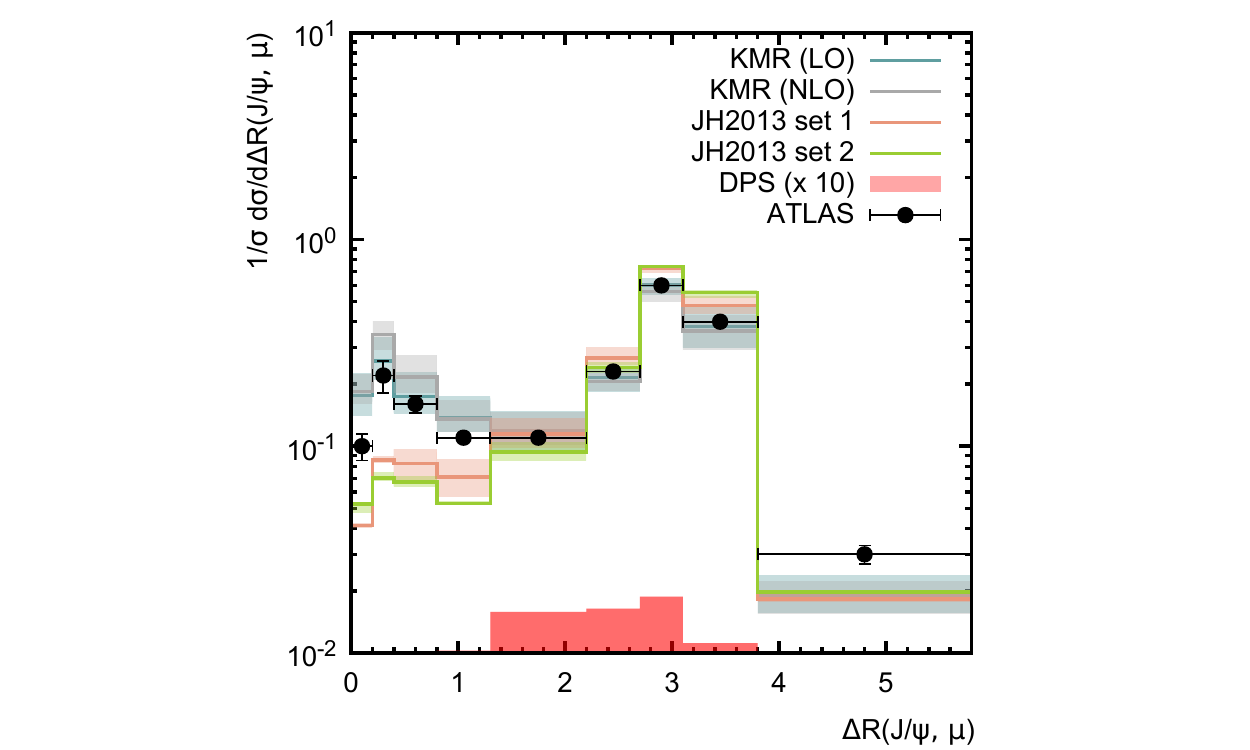}
\includegraphics[width=7.6cm]{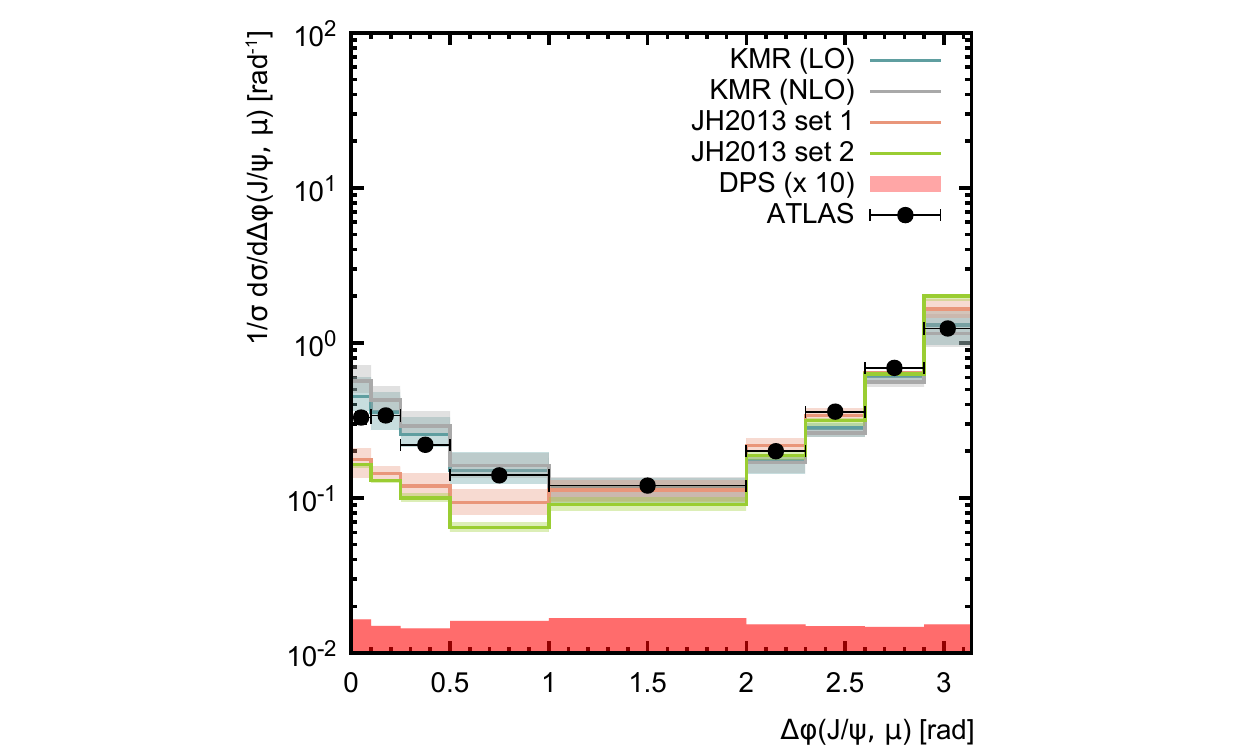}
\includegraphics[width=7.6cm]{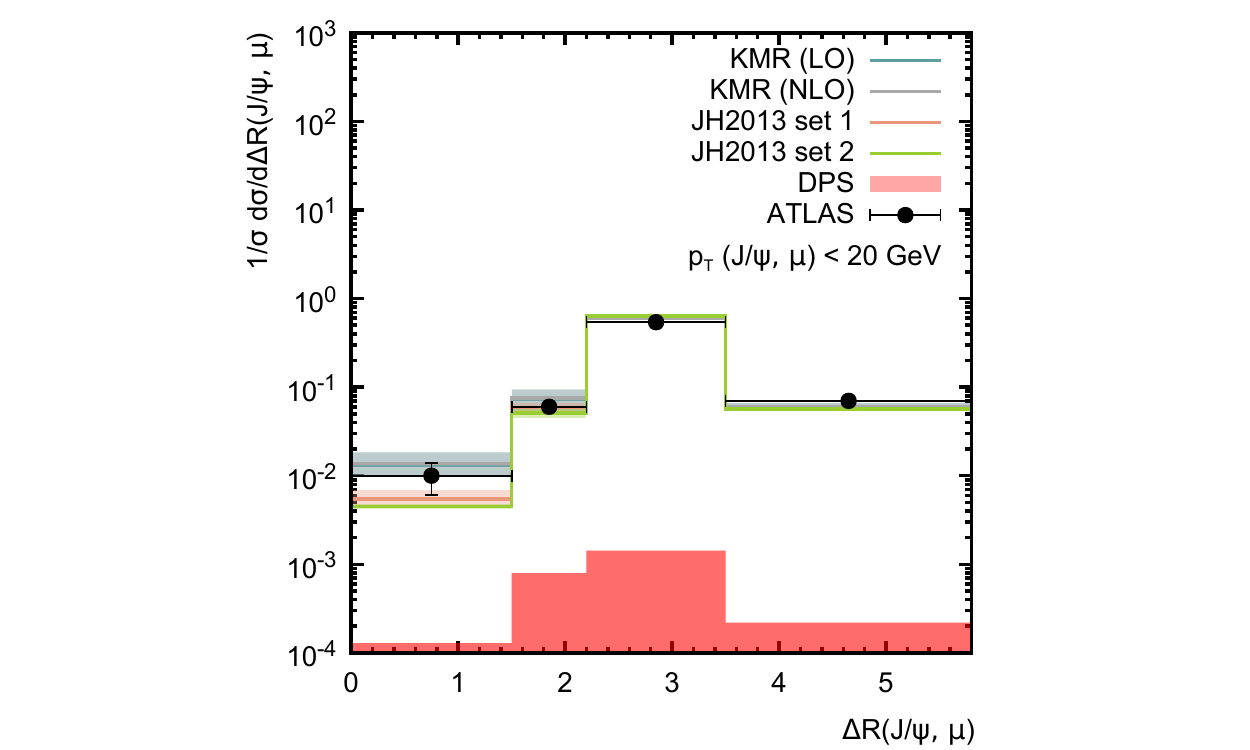}
\includegraphics[width=7.6cm]{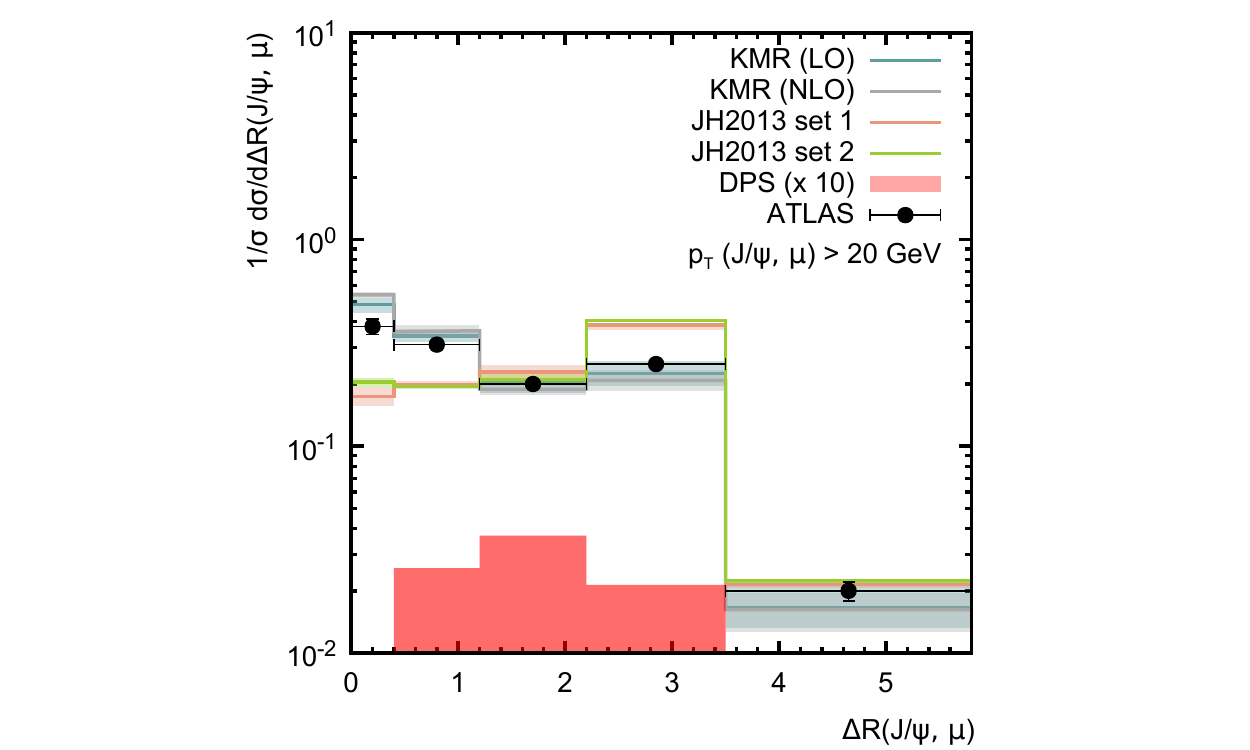}
\caption{The normalized differential cross sections of associated non-prompt $J/\psi + \mu$ 
production at $\sqrt s = 8$~TeV as a function of $\Delta R(J/\psi, \mu)$, $\Delta \phi(J/\psi, \mu)$,
low and high-$p_T$~$\Delta R(J/\psi, \mu)$. Predictions are made using the KMR (calculated 
with the LO and NLO accuracy) and CCFM-evolved TMD gluon densities. 
The shaded bands represent the scale uncertainties of the calculations, as it is 
described in the text. The DPS contributions are estimated using {\sl JH'2013 set 2} gluon 
density. The experimental data are from ATLAS\cite{1}.}
\label{fig1}
\end{center}
\end{figure}

\begin{figure}
\begin{center}
\includegraphics[width=7.6cm]{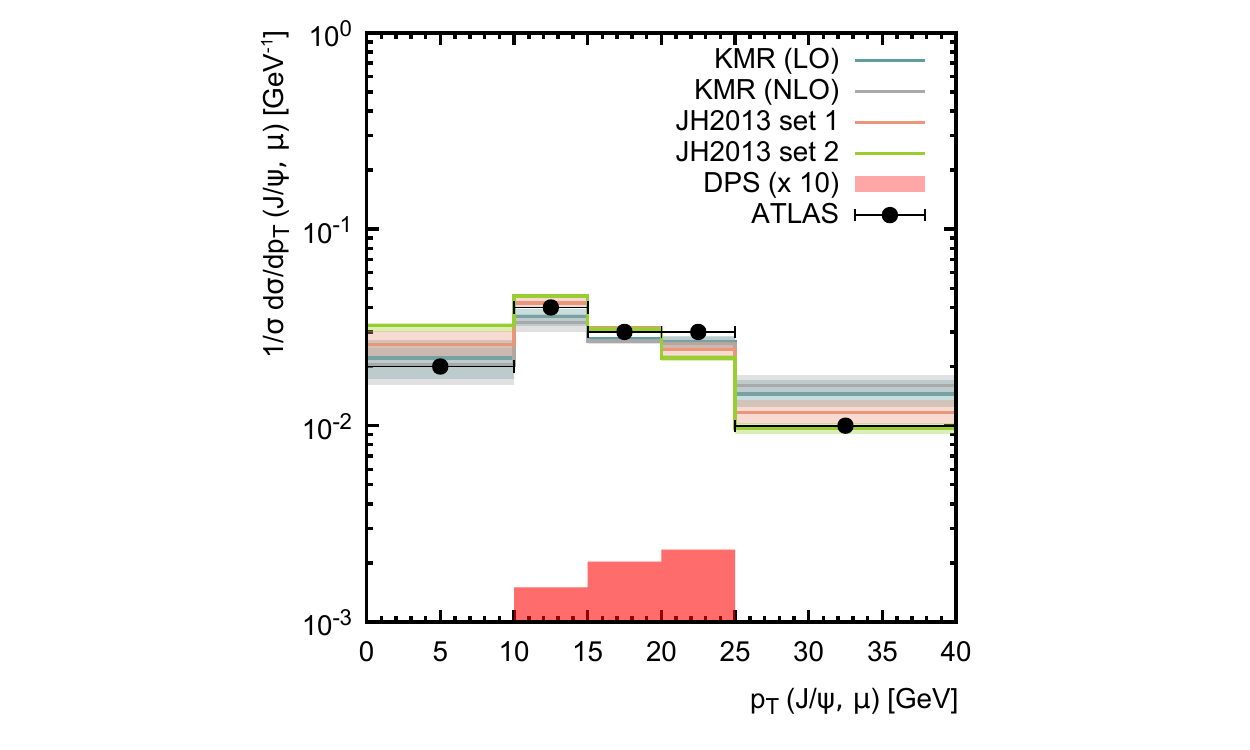}
\includegraphics[width=7.6cm]{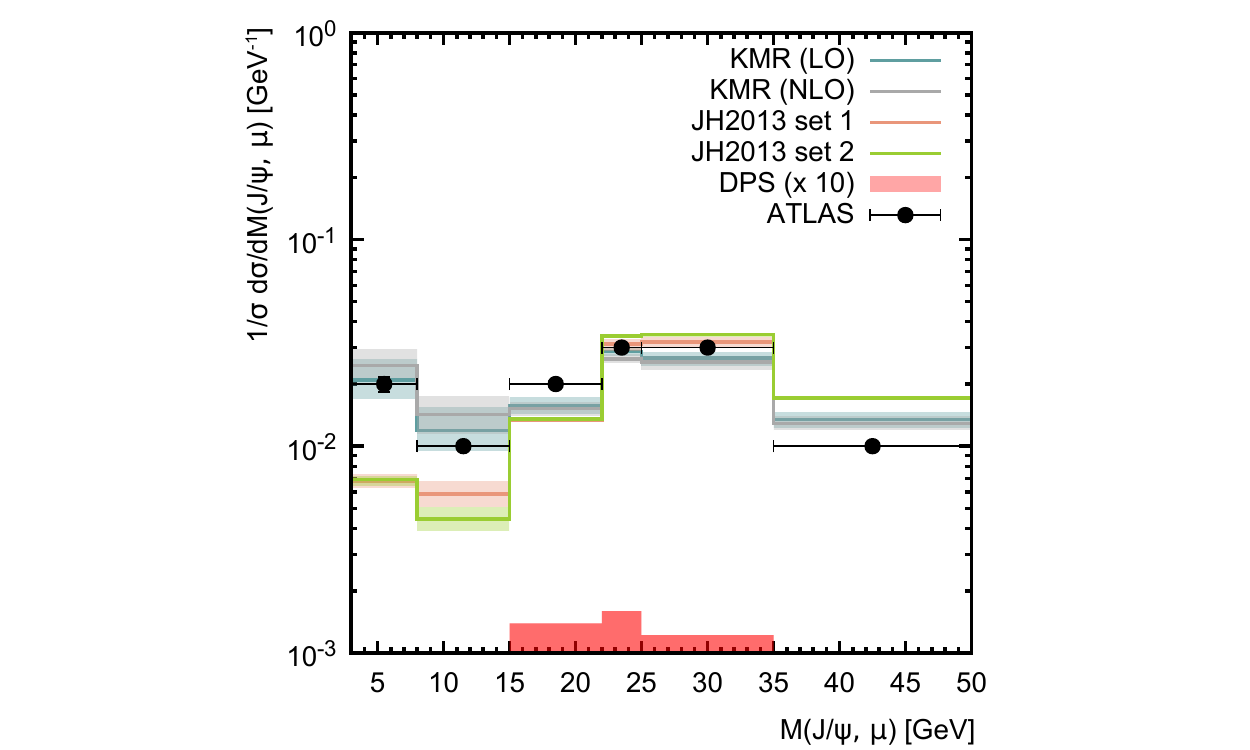}
\includegraphics[width=7.6cm]{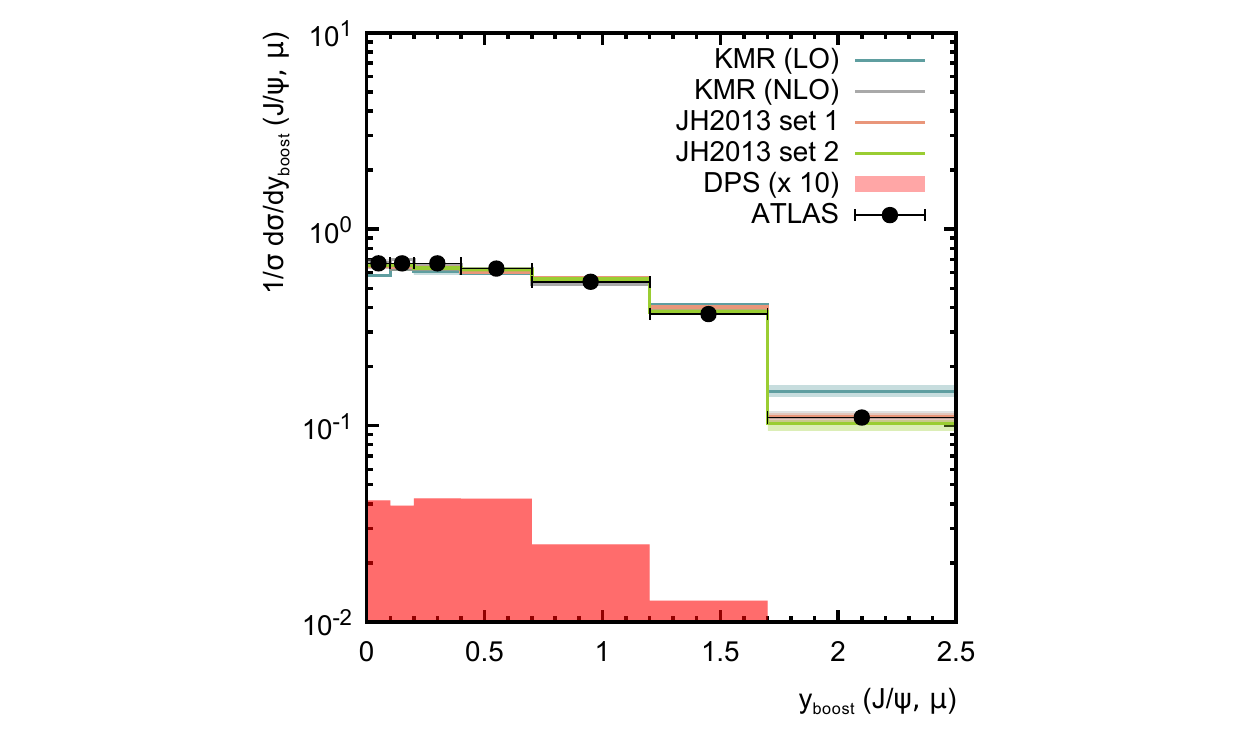}
\includegraphics[width=7.6cm]{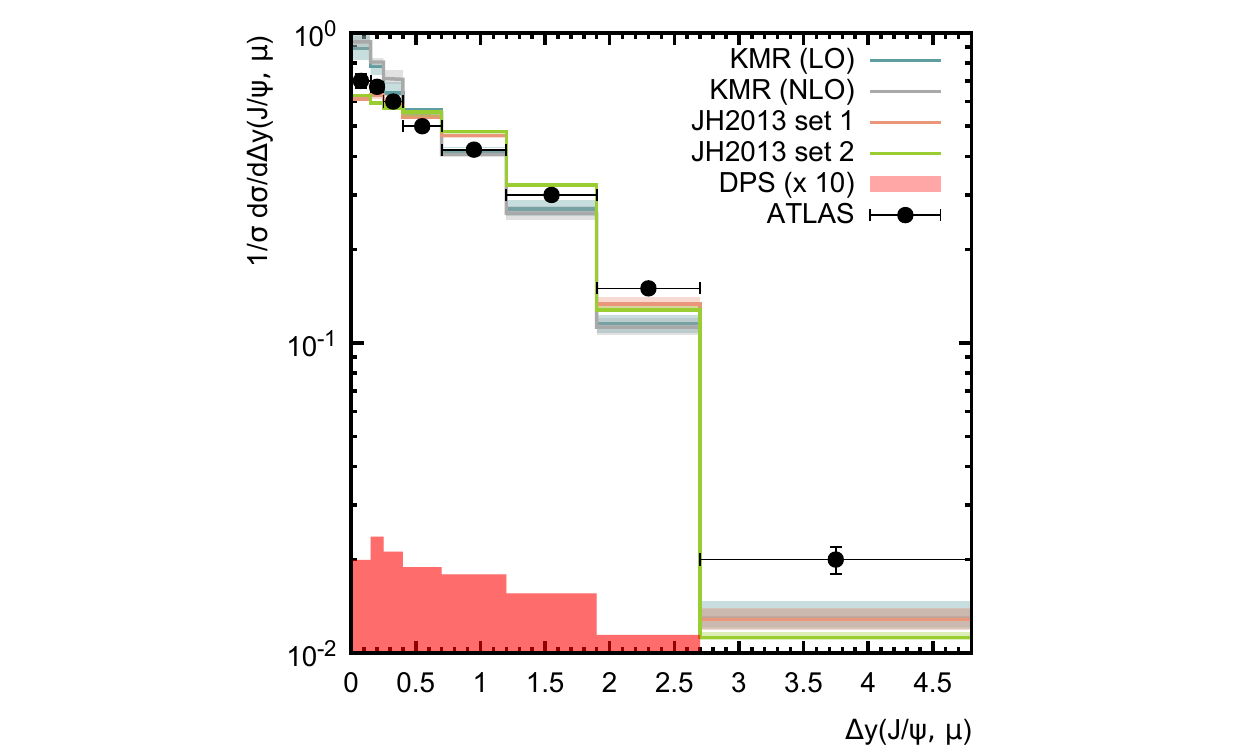}
\caption{The normalized differential cross sections of associated non-prompt $J/\psi + \mu$ 
production at $\sqrt s = 8$~TeV as a function of $p_T(J/\psi, \mu)$, $M(J/\psi, \mu)$, $y_{\rm boost}(J/\psi, \mu)$
and $\Delta y(J/\psi, \mu)$. Notation of histograms 
is the same as in Fig.~1. The experimental data are from ATLAS\cite{1}.}
\label{fig2}
\end{center}
\end{figure}

\begin{figure}
\begin{center}
\includegraphics[width=7.6cm]{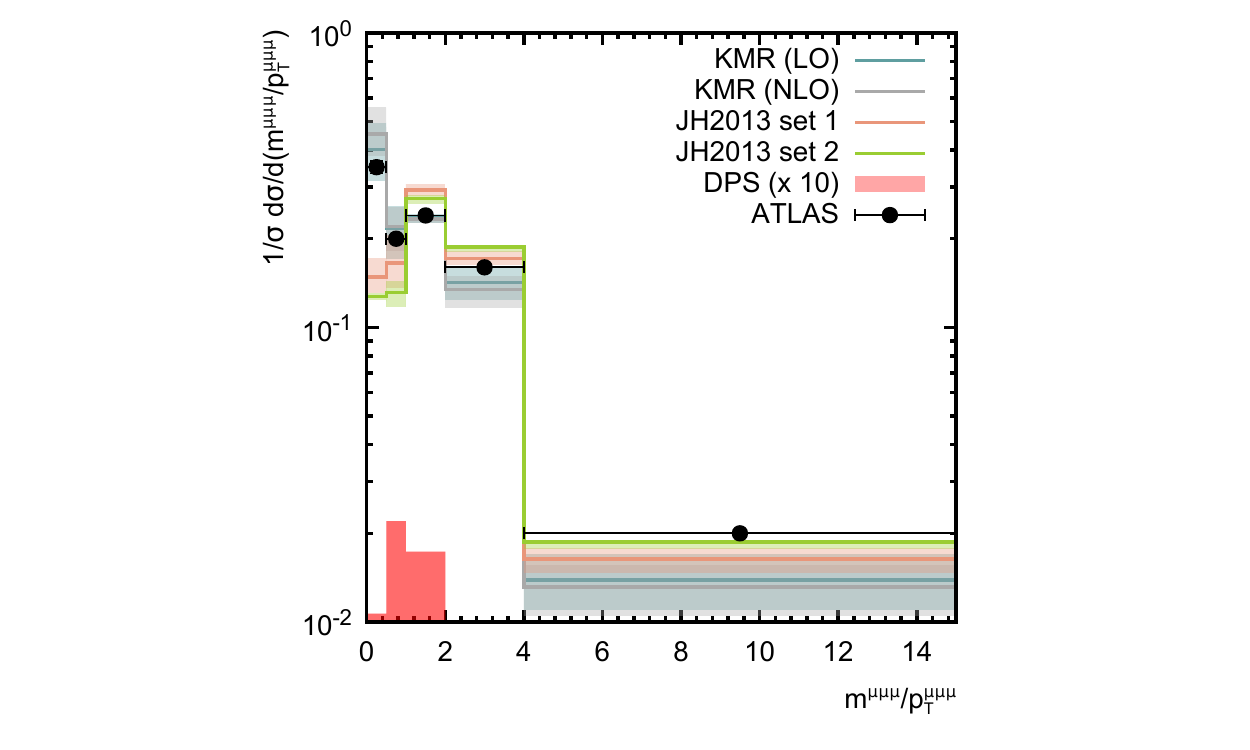}
\includegraphics[width=7.6cm]{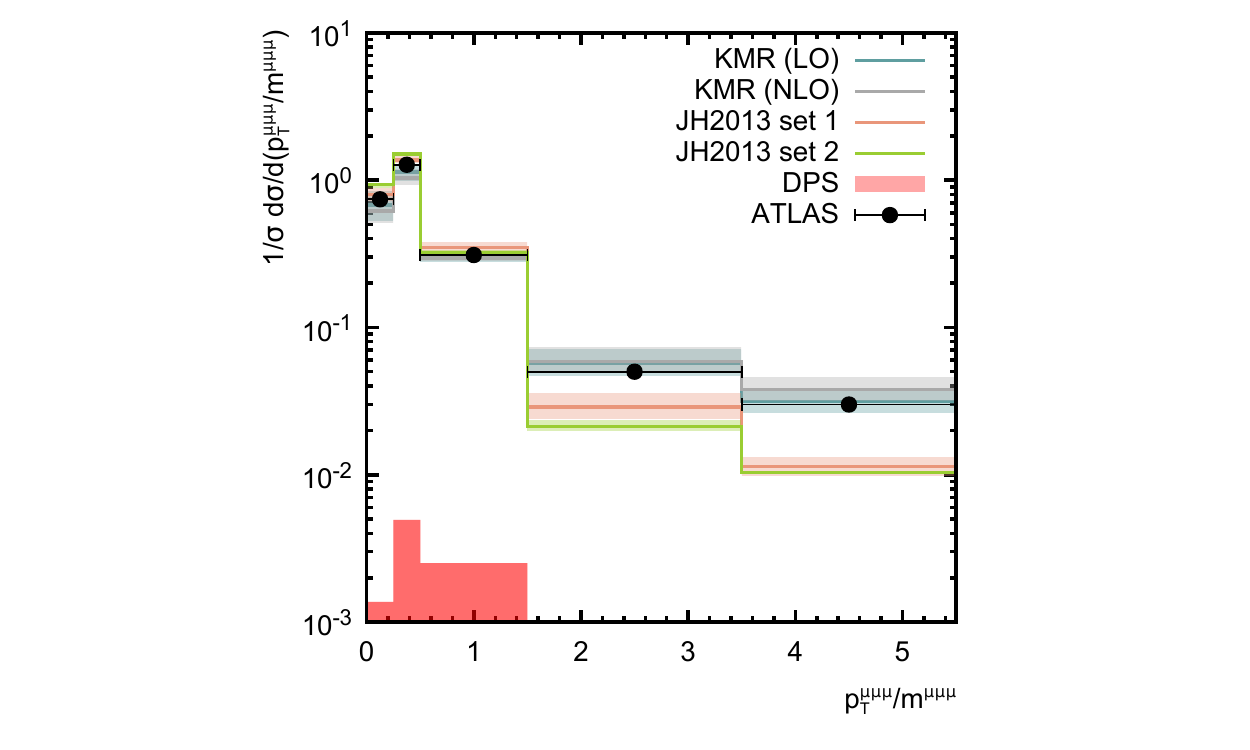}
\caption{The normalized differential cross sections of associated non-prompt $J/\psi + \mu$ 
production at $\sqrt s = 8$~TeV as a function of $m^{\mu\mu\mu}/p_T^{\mu\mu\mu}$ and $p_T^{\mu\mu\mu}/m^{\mu\mu\mu}$.
Notation of histograms is the same as in Fig.~1. The experimental data are from ATLAS\cite{1}.}
\label{fig3}
\end{center}
\end{figure}

\begin{figure}
\begin{center}
\includegraphics[width=7.6cm]{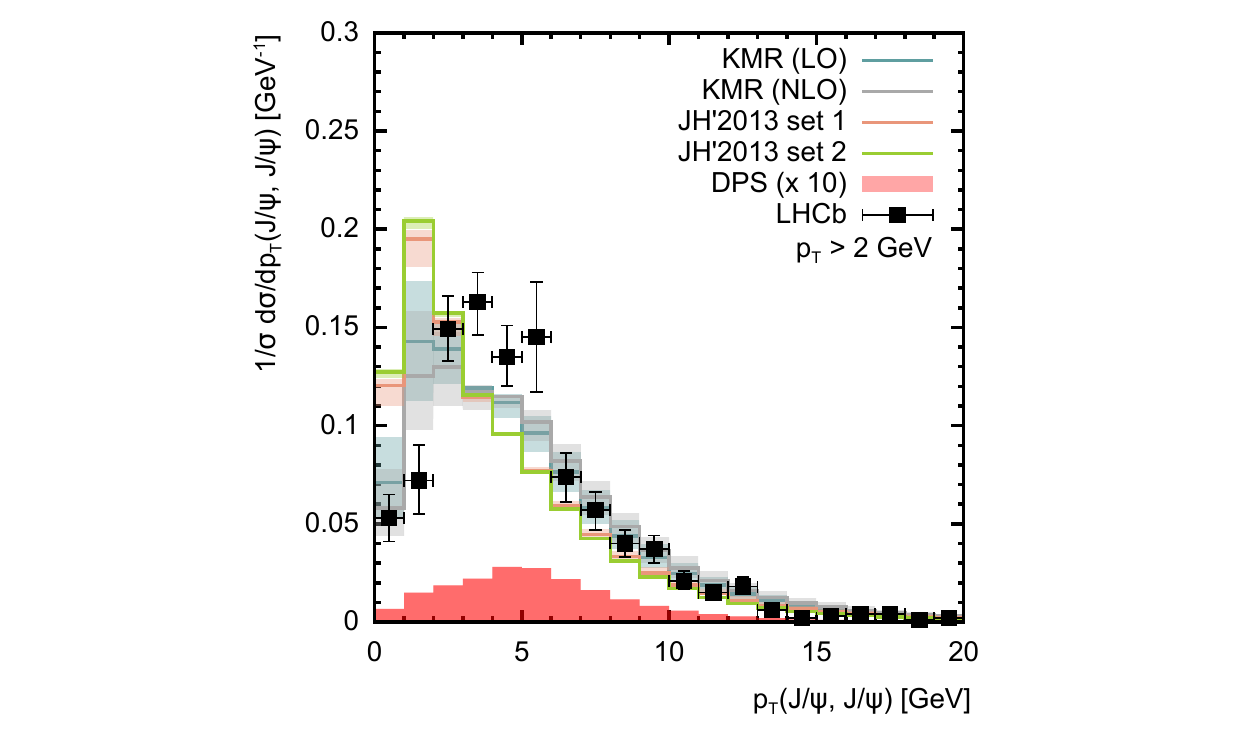}
\includegraphics[width=7.6cm]{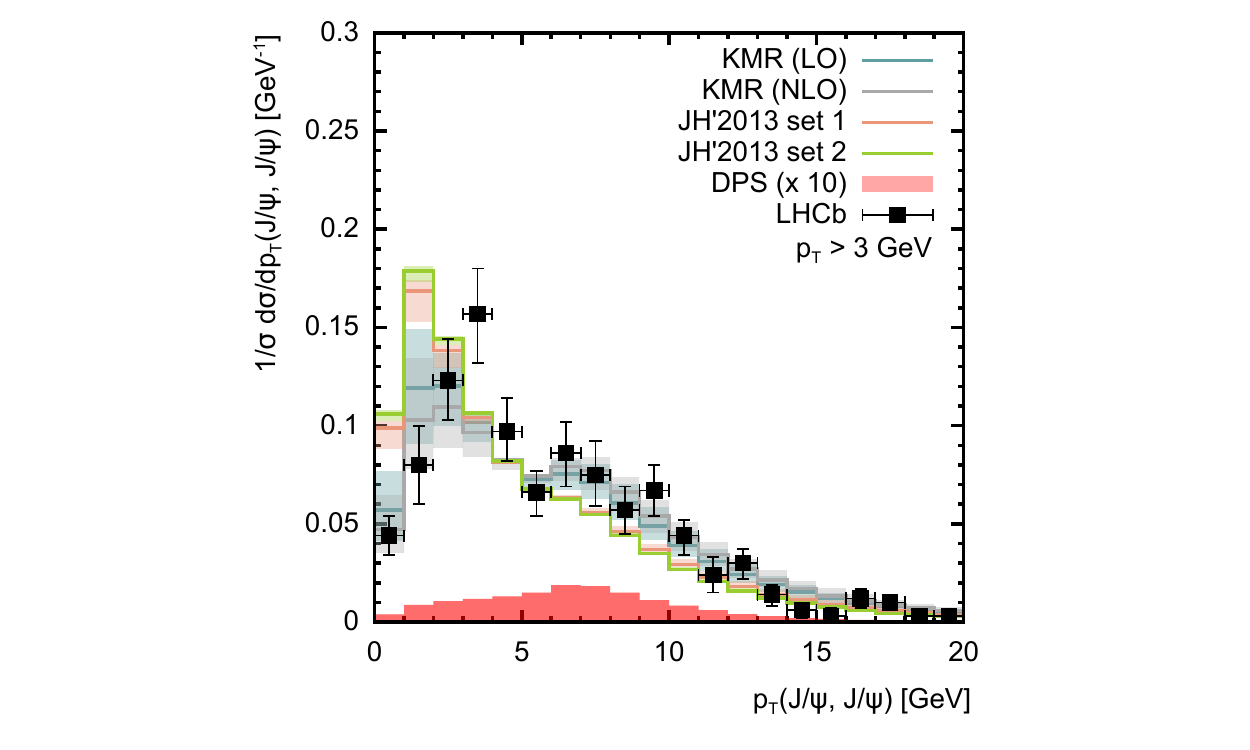}
\includegraphics[width=7.6cm]{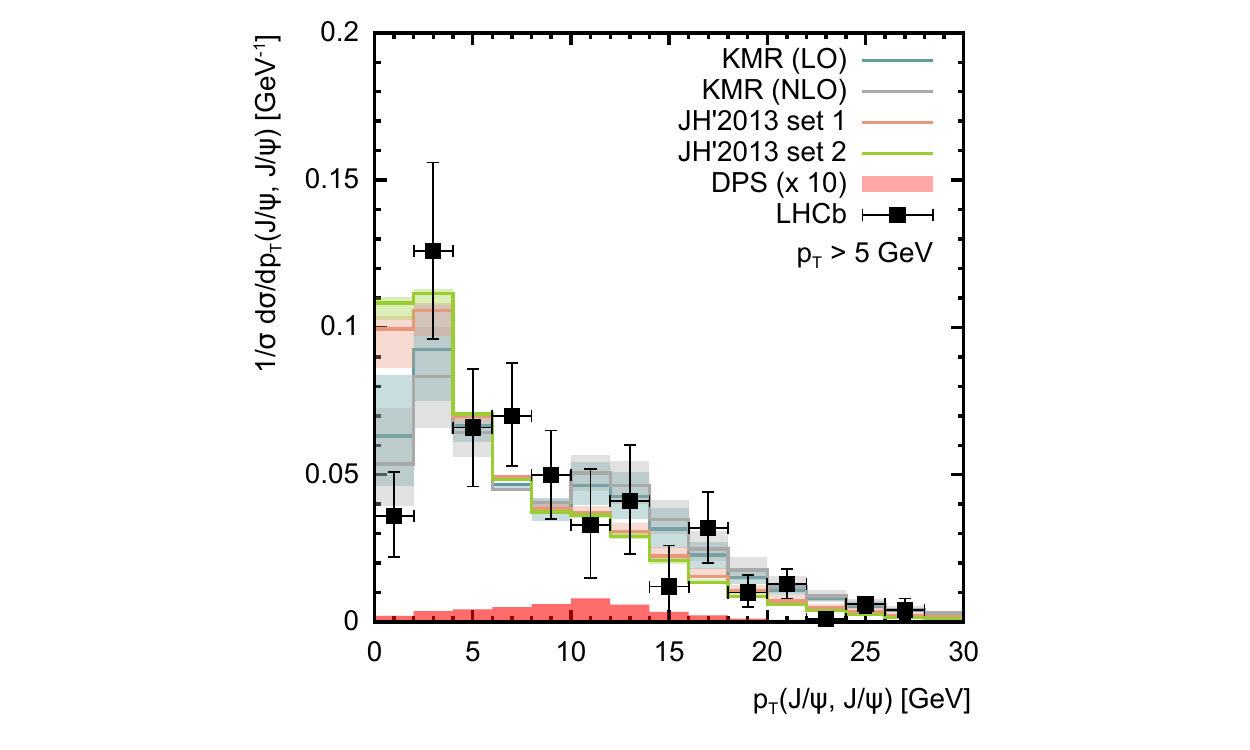}
\includegraphics[width=7.6cm]{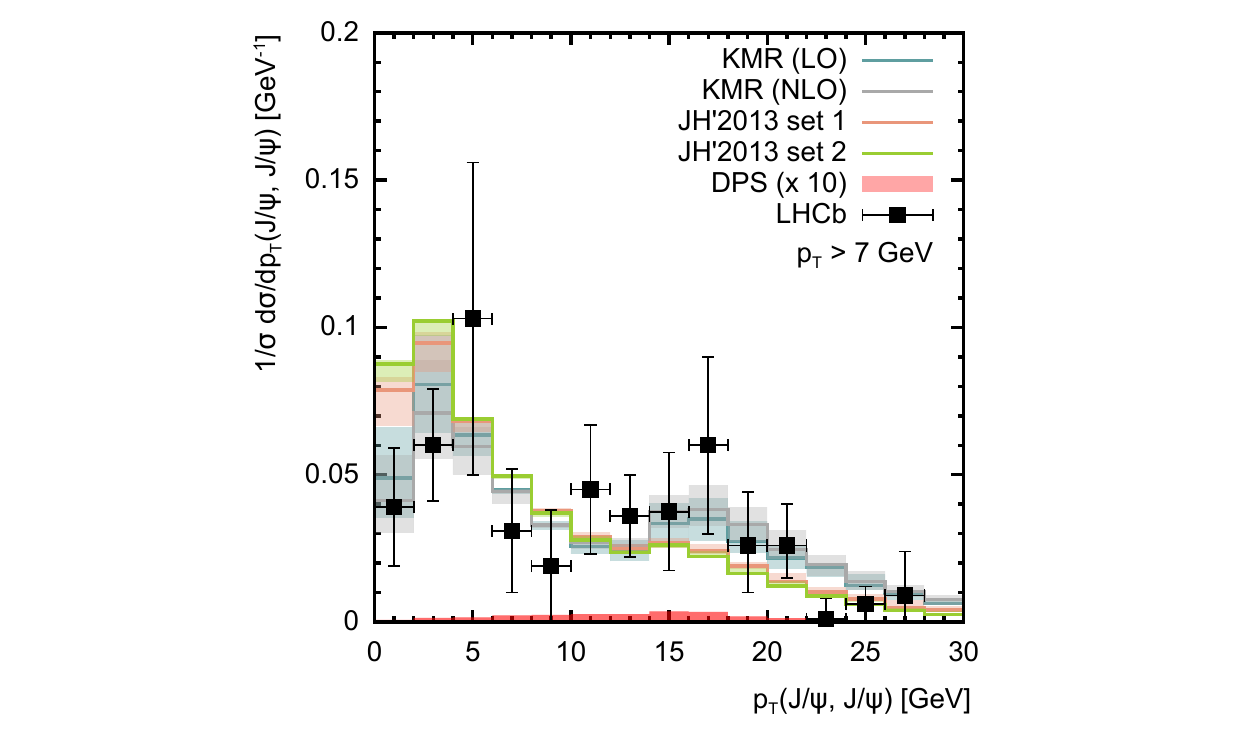}
\caption{The normalized differential cross sections of non-prompt $J/\psi + J/\psi$ 
production at $\sqrt s = 8$~TeV as a function of the transverse momentum of the $J/\psi$ pair.
Notation of histograms is the same as in Fig.~1. The experimental data are from LHCb\cite{2}.}
\label{fig4}
\end{center}
\end{figure}

\begin{figure}
\begin{center}
\includegraphics[width=7.6cm]{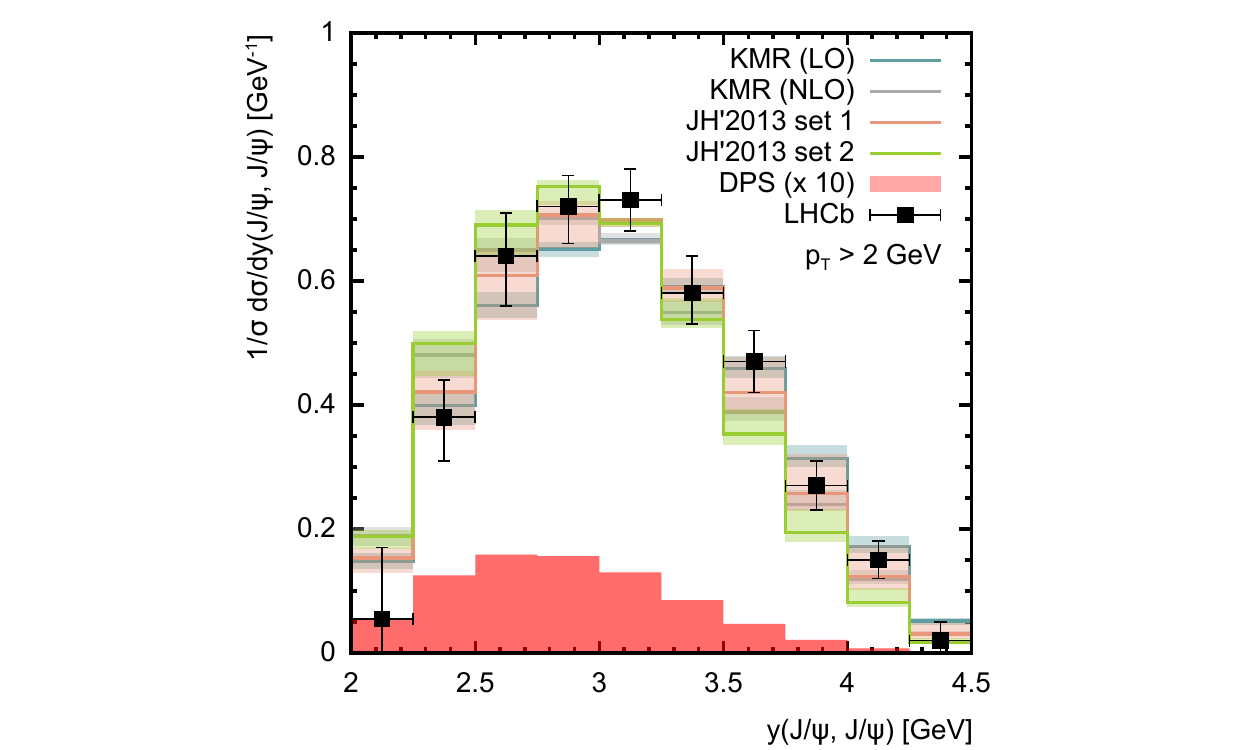}
\includegraphics[width=7.6cm]{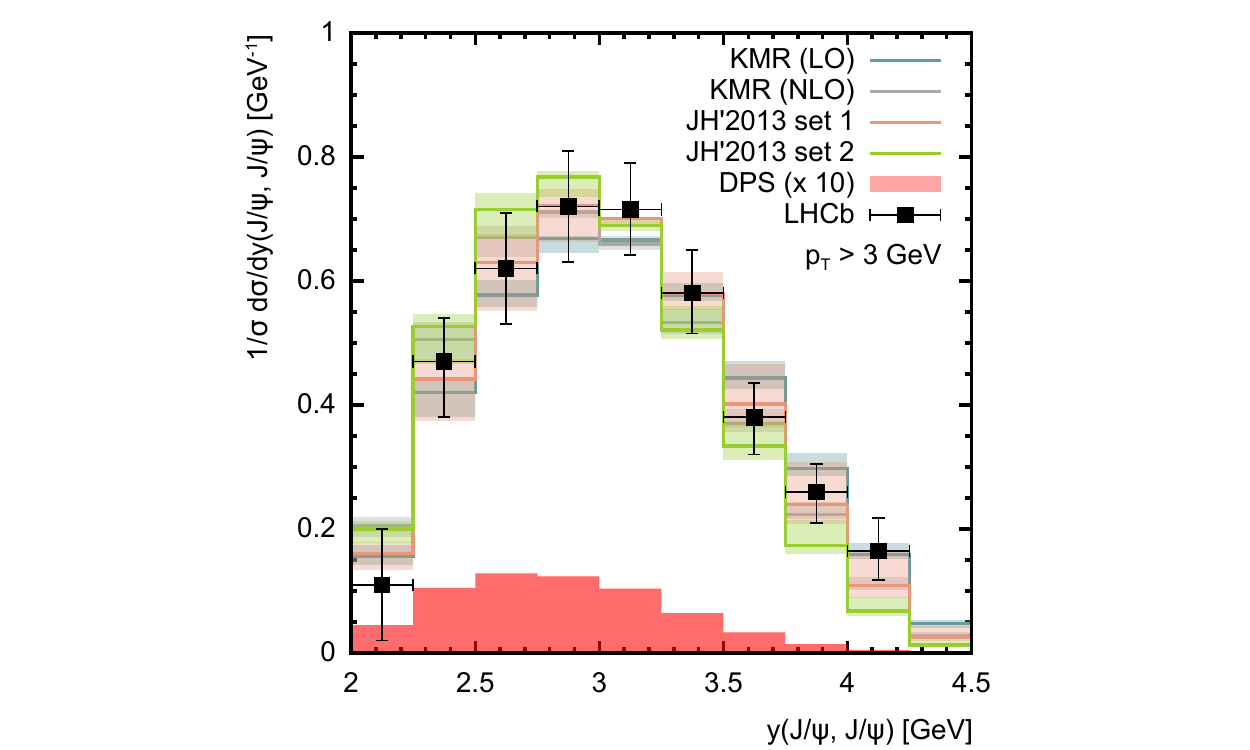}
\includegraphics[width=7.6cm]{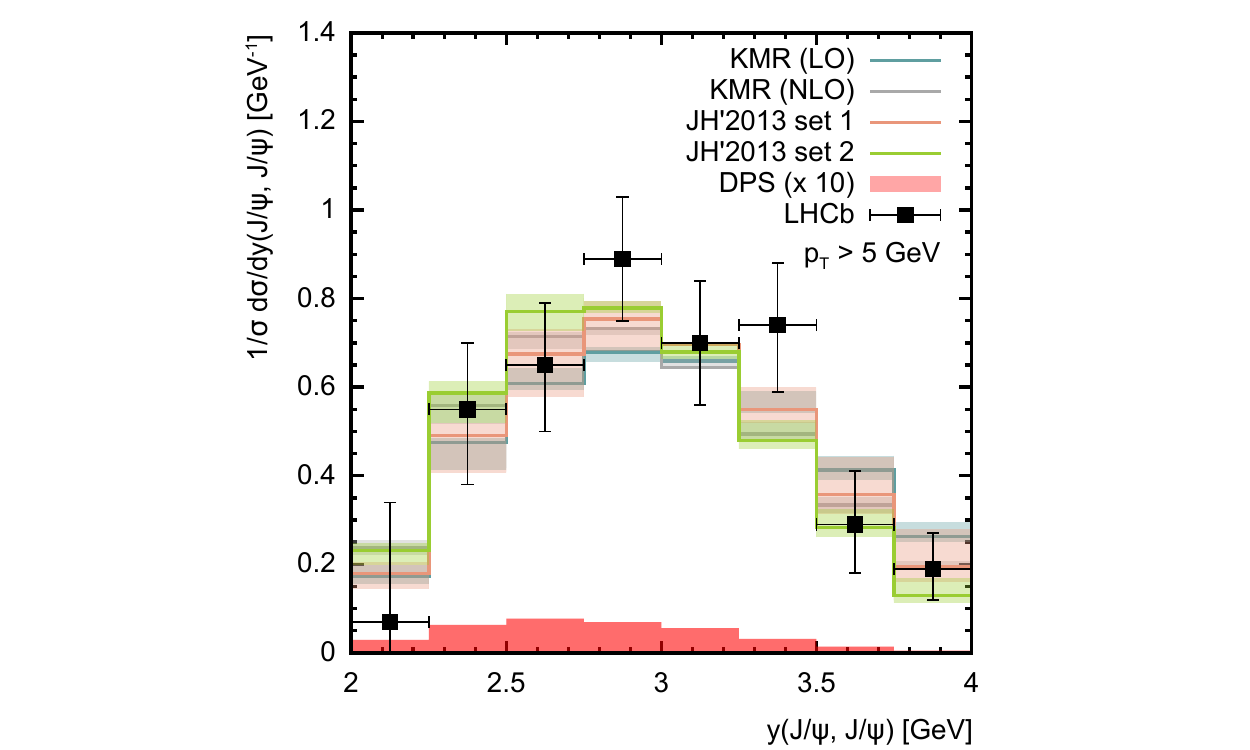}
\includegraphics[width=7.6cm]{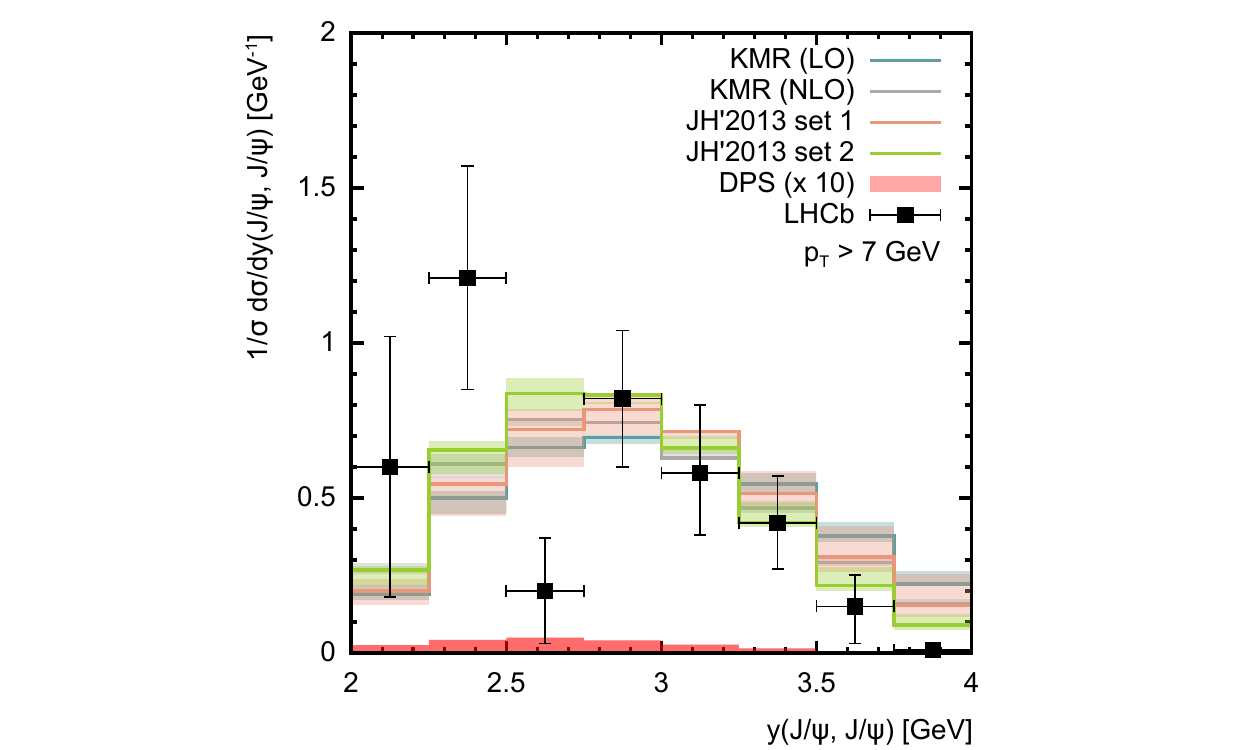}
\caption{The normalized differential cross sections of non-prompt $J/\psi + J/\psi$ 
production at $\sqrt s = 8$~TeV as a function of the rapidity of the $J/\psi$ pair.
Notation of histograms is the same as in Fig.~1. The experimental data are from LHCb\cite{2}.}
\label{fig5}
\end{center}
\end{figure}

\begin{figure}
\begin{center}
\includegraphics[width=7.6cm]{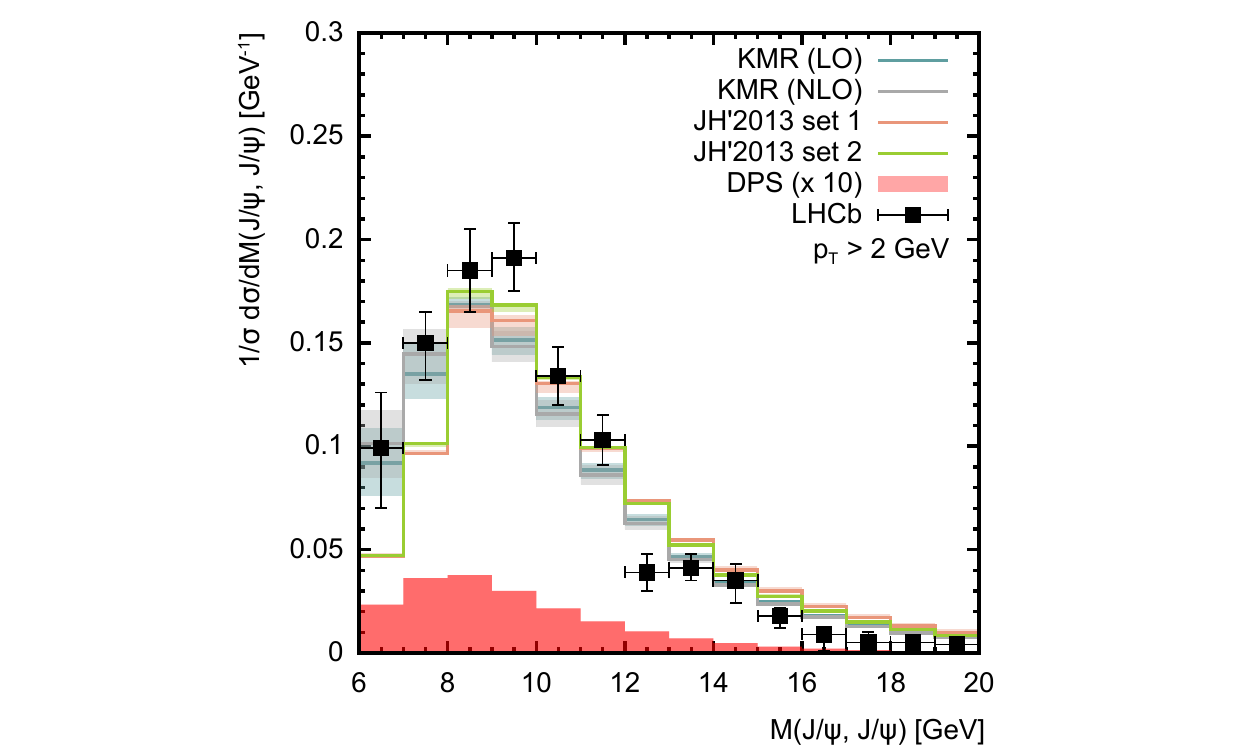}
\includegraphics[width=7.6cm]{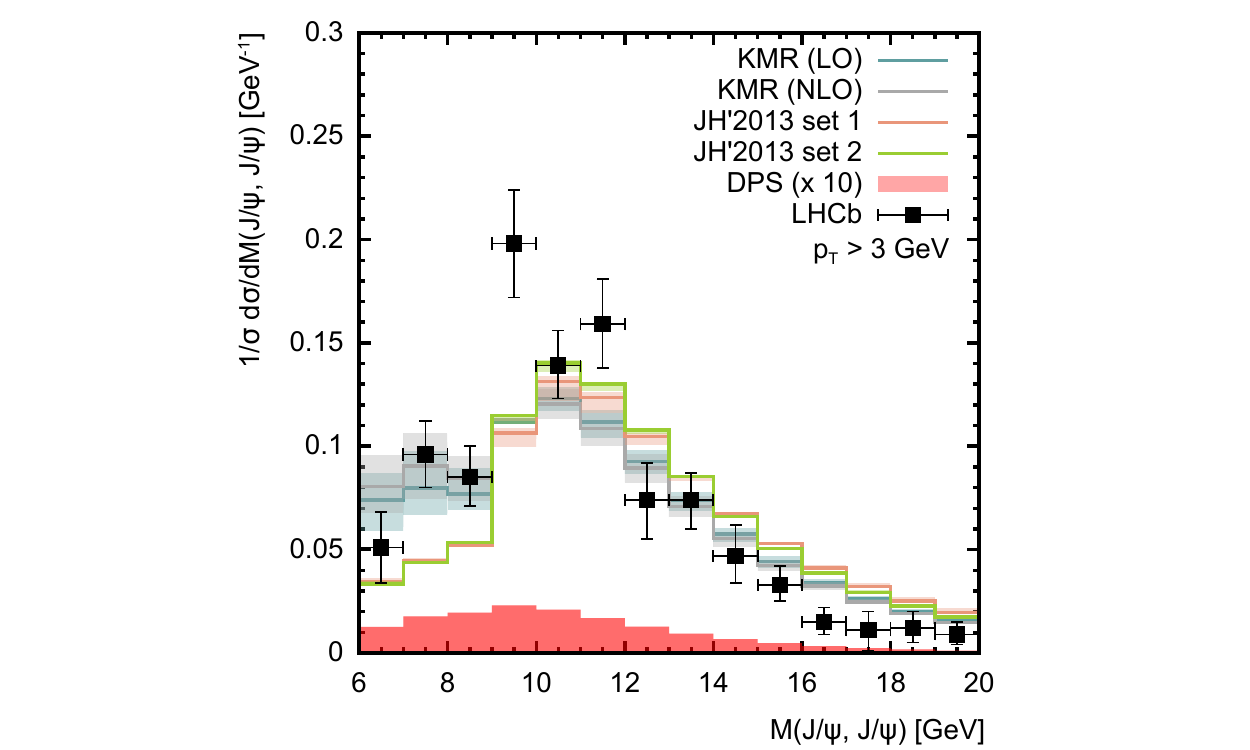}
\includegraphics[width=7.6cm]{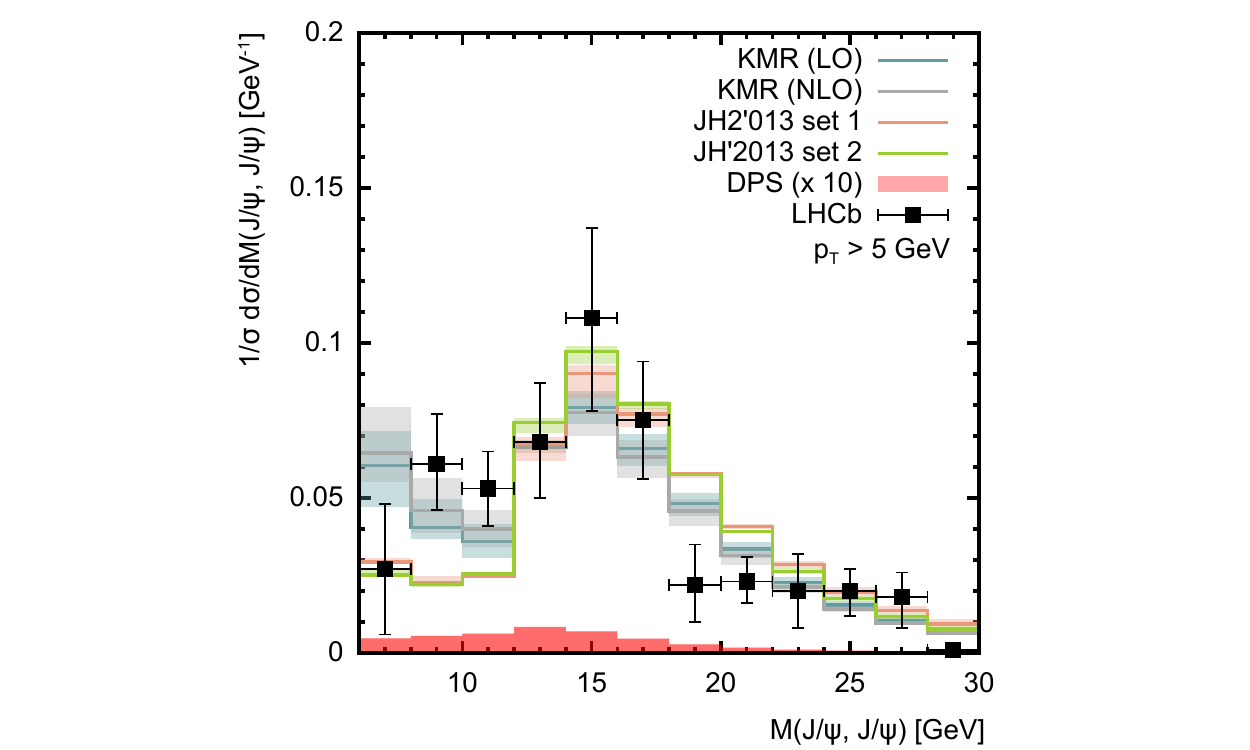}
\includegraphics[width=7.6cm]{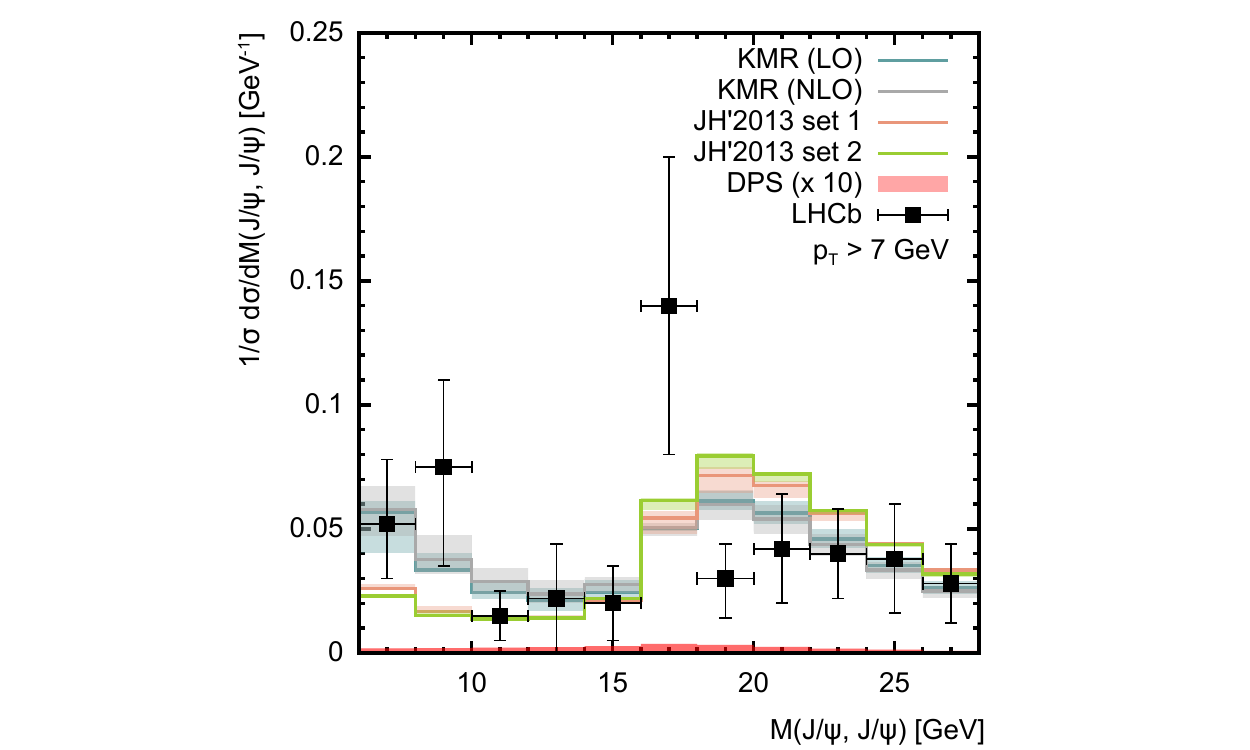}
\caption{The normalized differential cross sections of non-prompt $J/\psi + J/\psi$ 
production at $\sqrt s = 8$~TeV as a function of the invariant mass of the $J/\psi$ pair.
Notation of histograms is the same as in Fig.~1. The experimental data are from LHCb\cite{2}.}
\label{fig6}
\end{center}
\end{figure}

\begin{figure}
\begin{center}
\includegraphics[width=7.6cm]{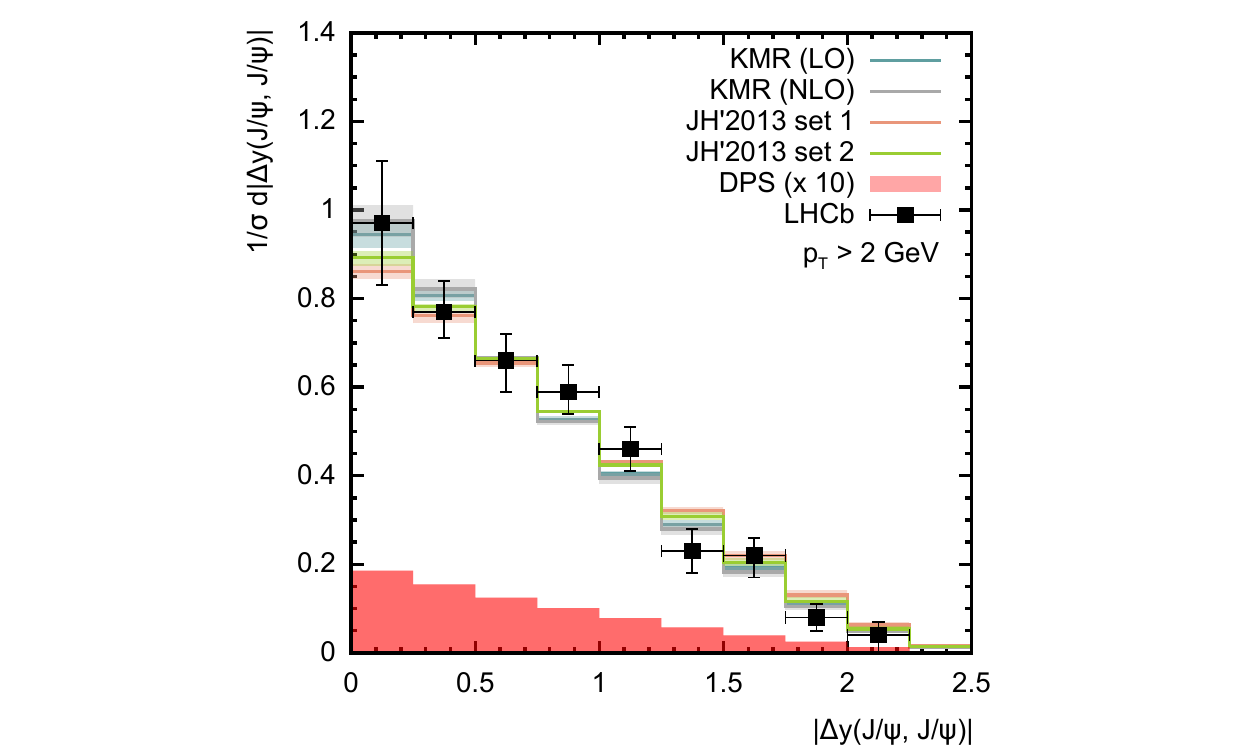}
\includegraphics[width=7.6cm]{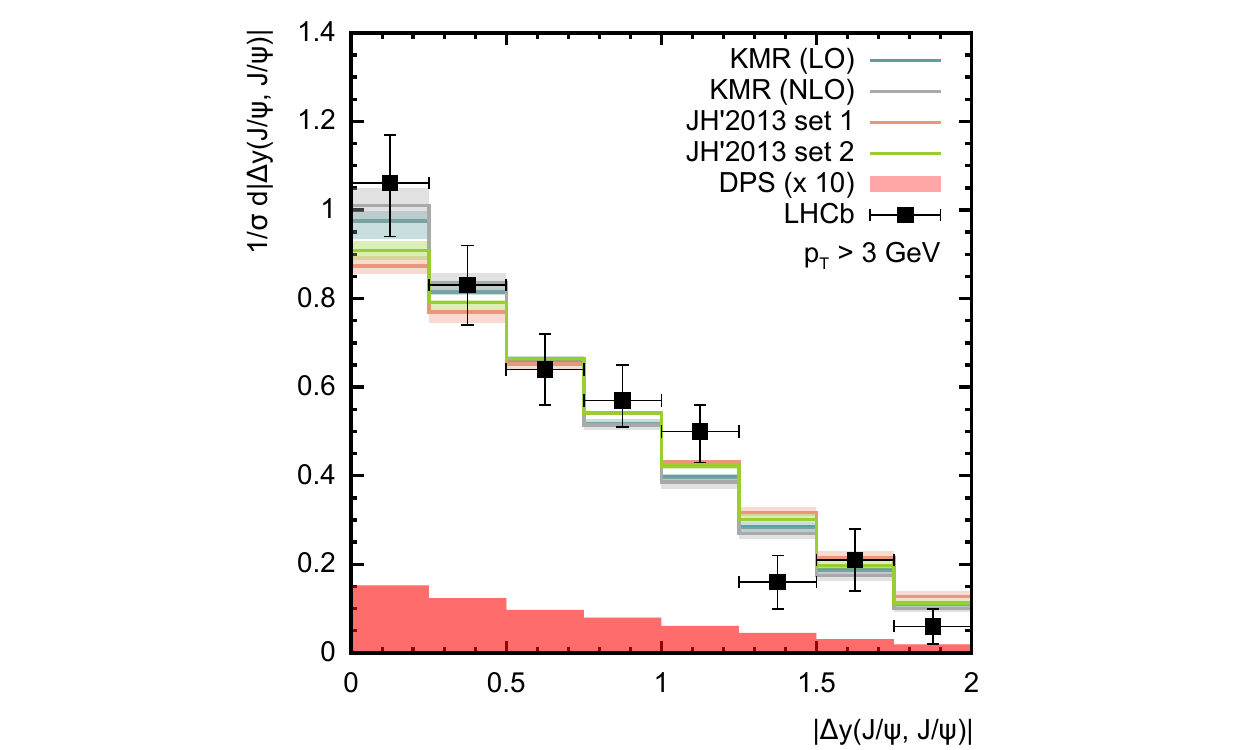}
\includegraphics[width=7.6cm]{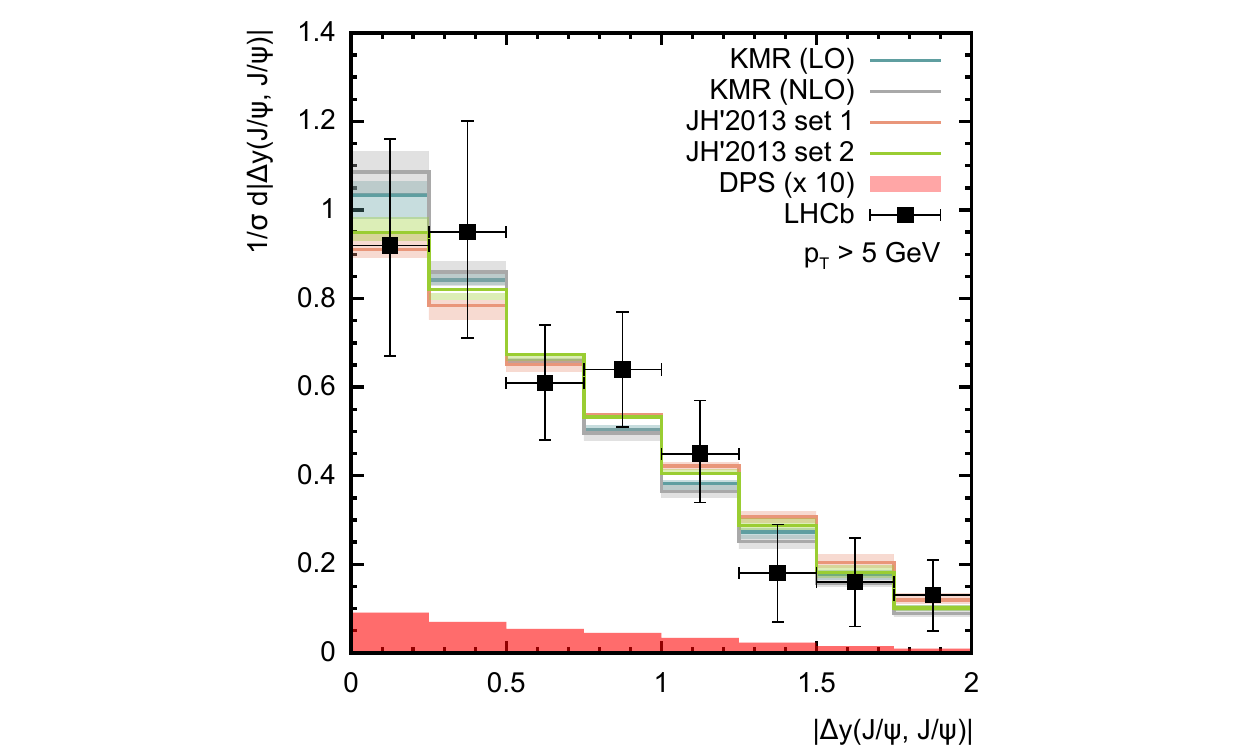}
\includegraphics[width=7.6cm]{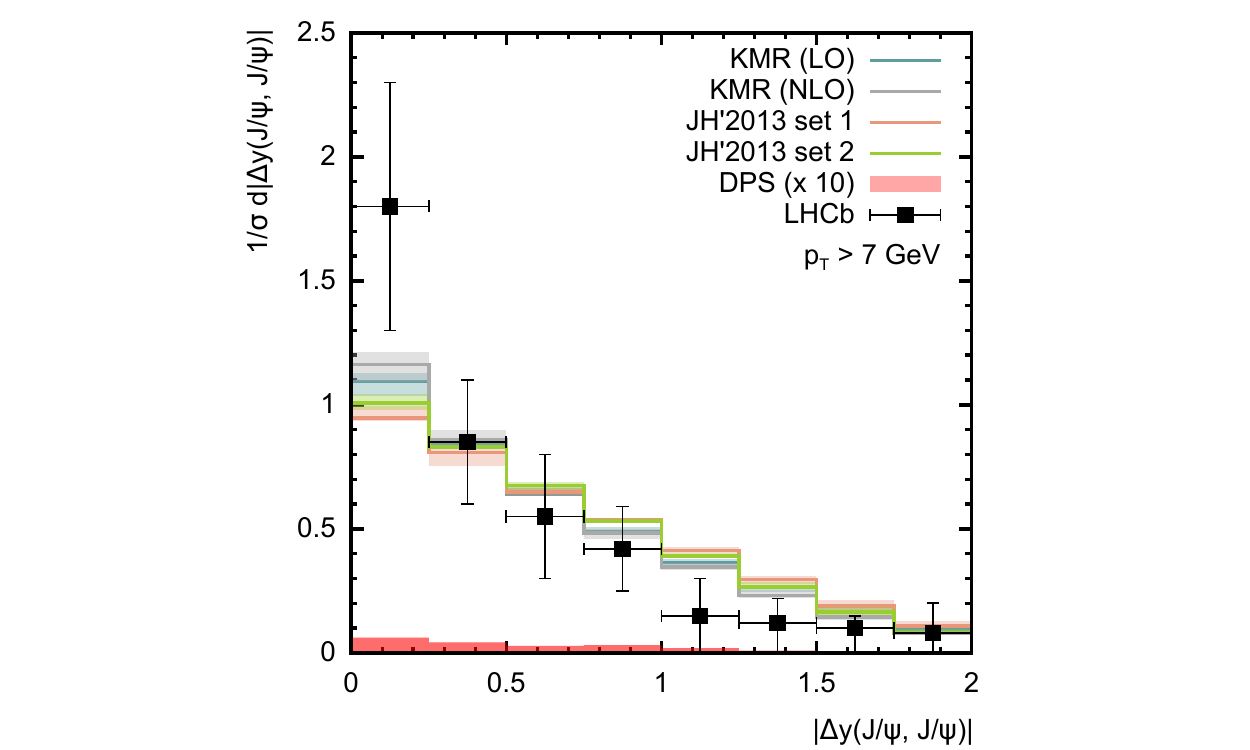}
\caption{The normalized differential cross sections of non-prompt $J/\psi + J/\psi$ 
production at $\sqrt s = 8$~TeV as a function of the difference in 
the rapidity between the two $J/\psi$ mesons.
Notation of histograms is the same as in Fig.~1. The experimental data are from LHCb\cite{2}.}
\label{fig7}
\end{center}
\end{figure}

\begin{figure}
\begin{center}
\includegraphics[width=7.6cm]{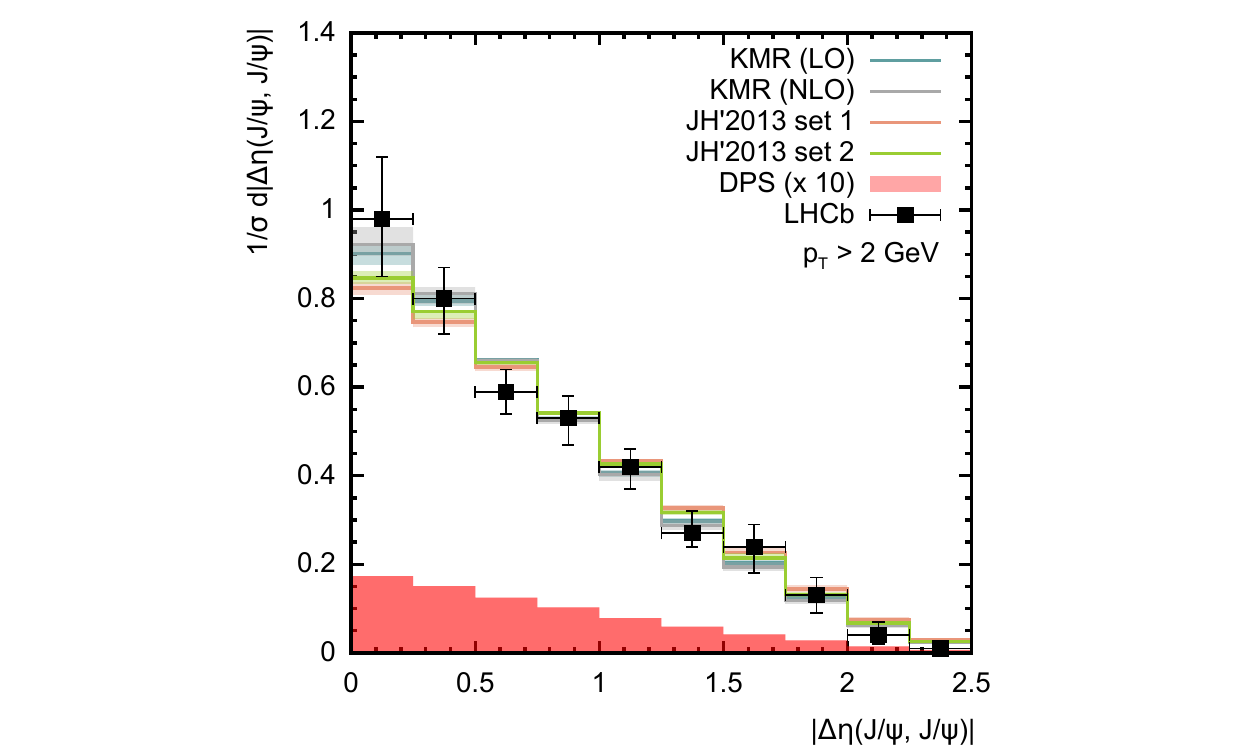}
\includegraphics[width=7.6cm]{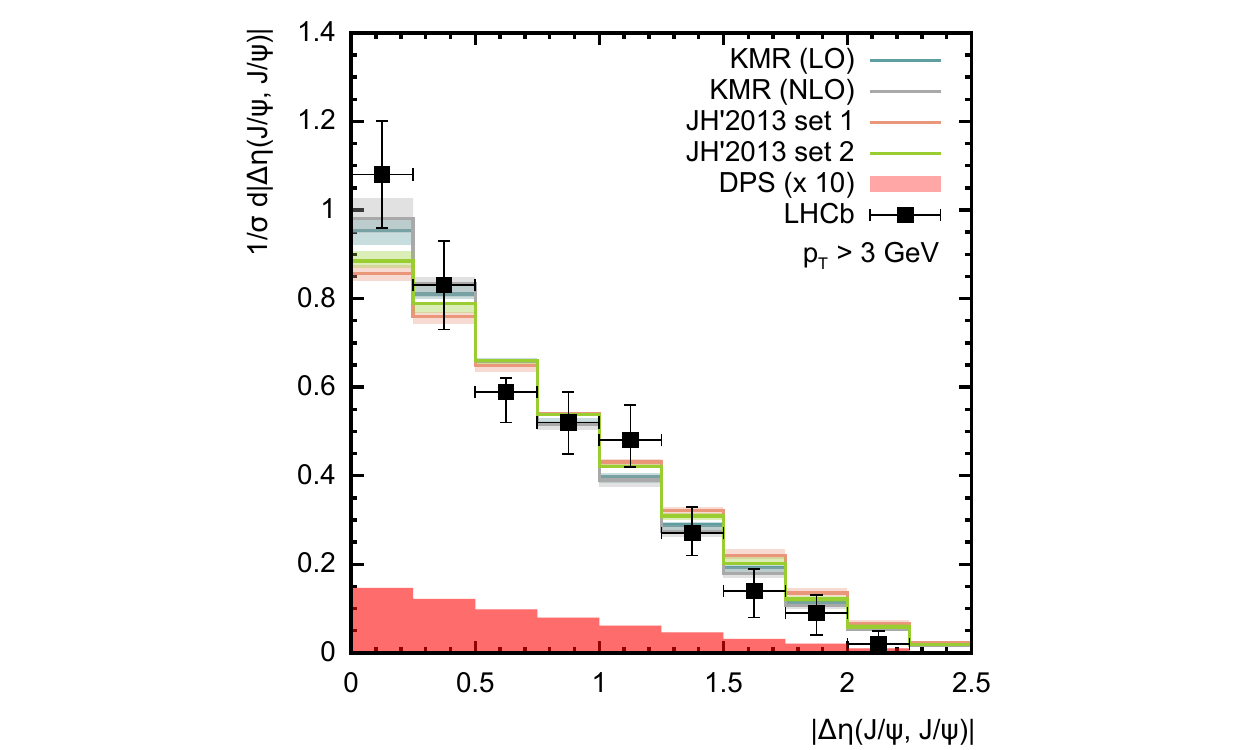}
\includegraphics[width=7.6cm]{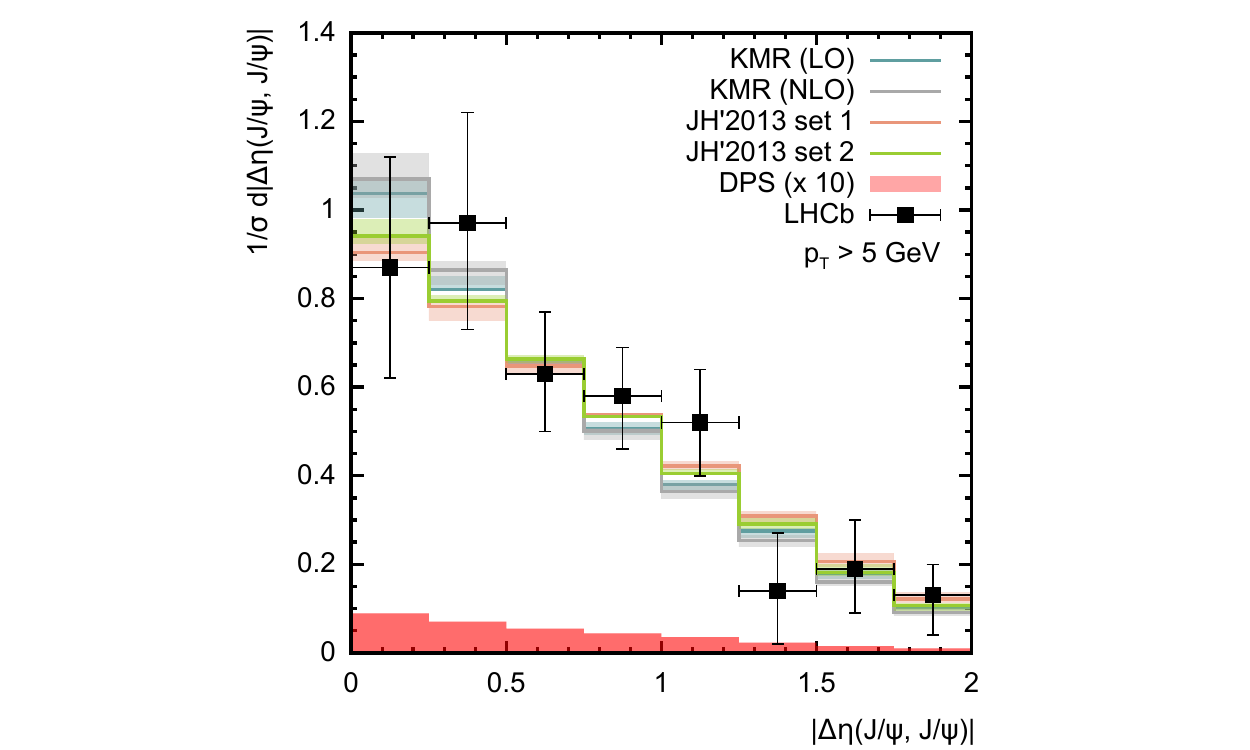}
\includegraphics[width=7.6cm]{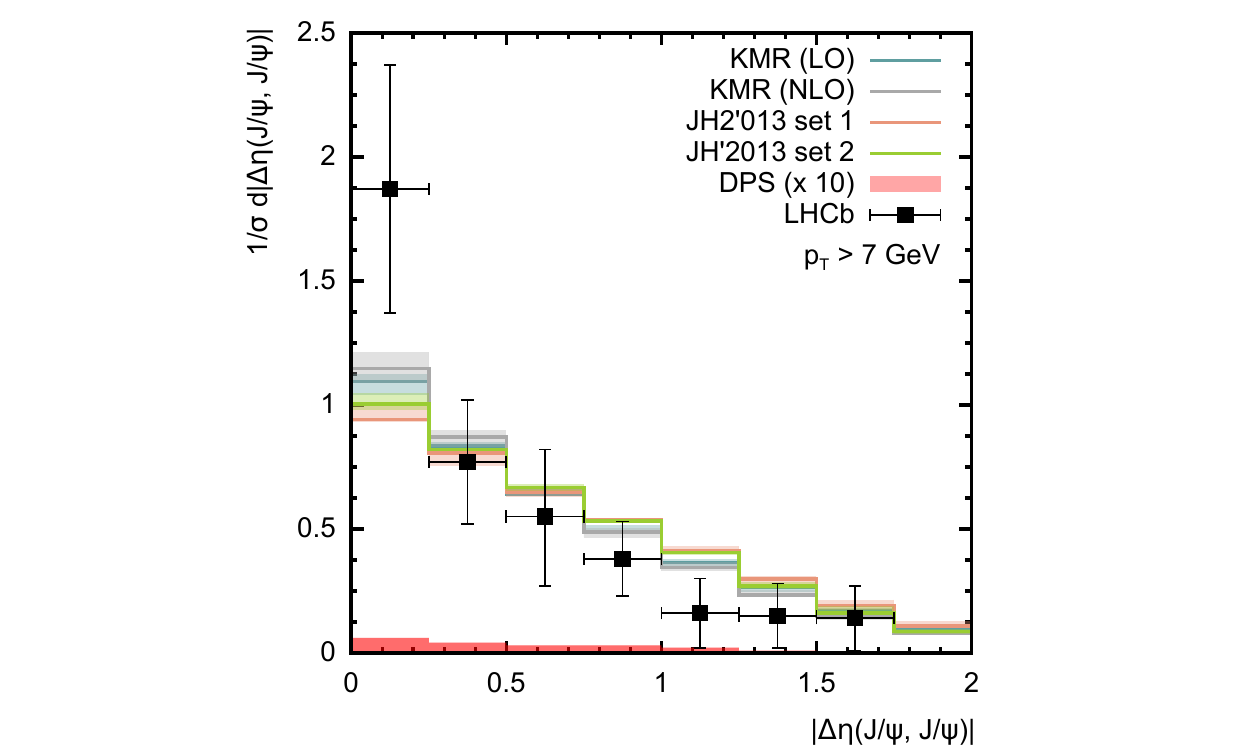}
\caption{The normalized differential cross sections of non-prompt $J/\psi + J/\psi$ 
production at $\sqrt s = 8$~TeV as a function of the difference in 
the pseudorapidity between the two $J/\psi$ mesons.
Notation of histograms is the same as in Fig.~1. The experimental data are from LHCb\cite{2}.}
\label{fig8}
\end{center}
\end{figure}

\begin{figure}
\begin{center}
\includegraphics[width=7.6cm]{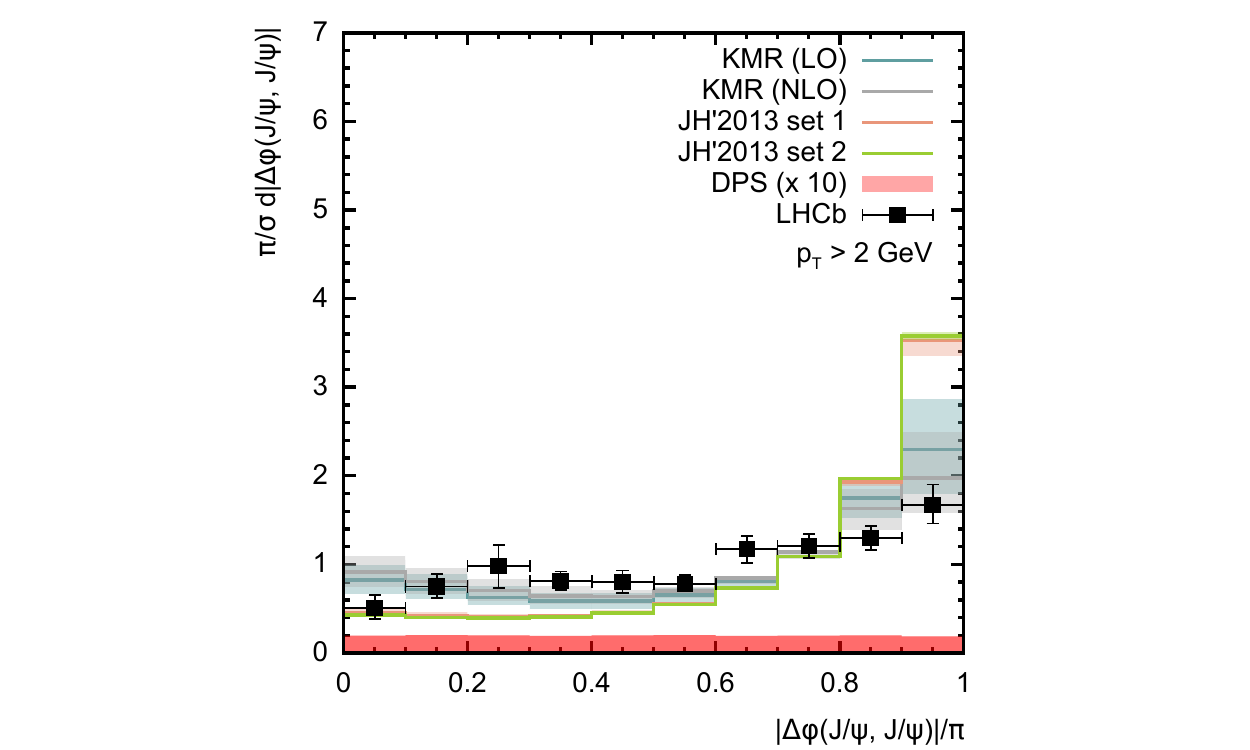}
\includegraphics[width=7.6cm]{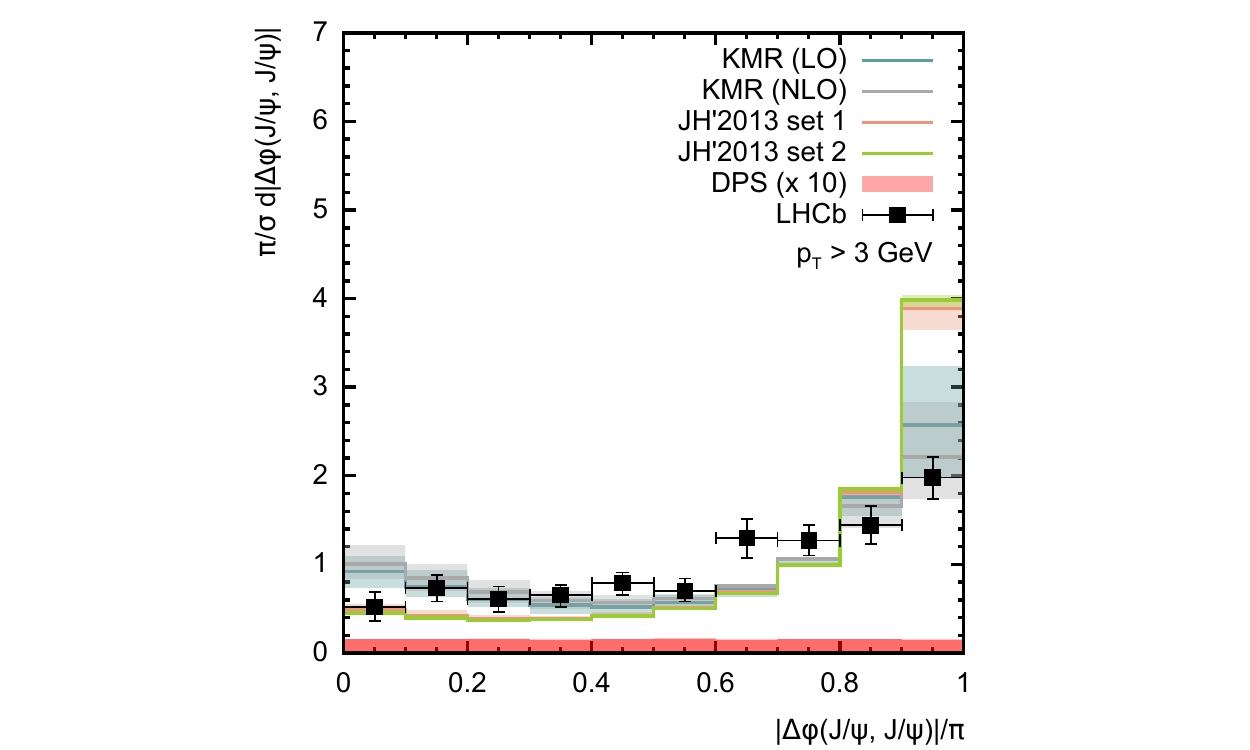}
\includegraphics[width=7.6cm]{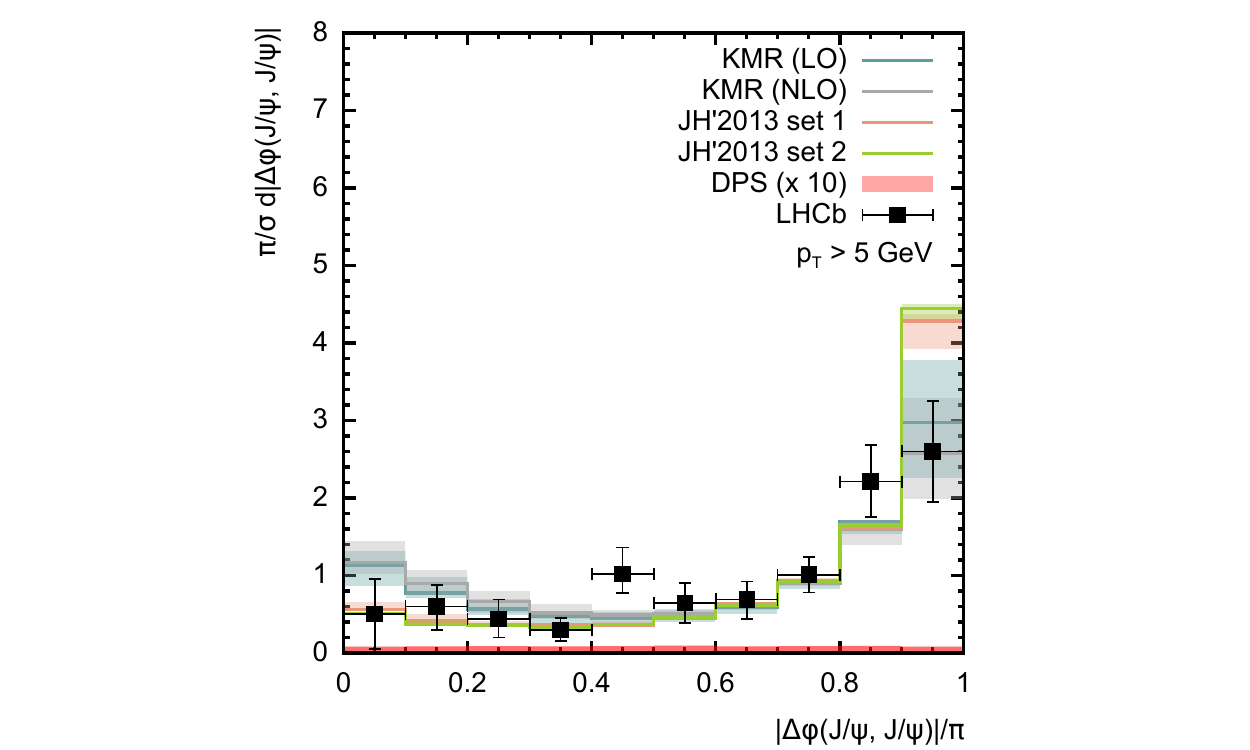}
\includegraphics[width=7.6cm]{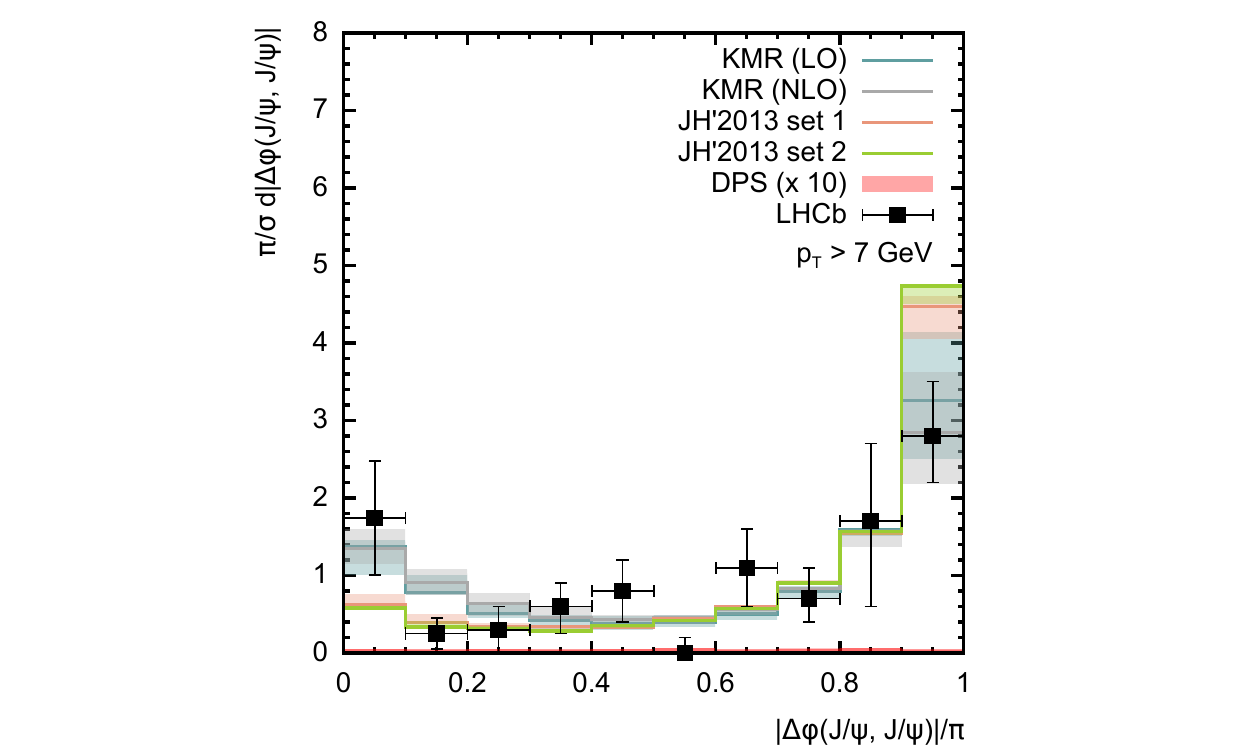}
\caption{The normalized differential cross sections of non-prompt $J/\psi + J/\psi$ 
production at $\sqrt s = 8$~TeV as a function of the difference in 
the azimuthal angle between the momentum directions of two $J/\psi$ mesons.
Notation of histograms is the same as in Fig.~1. The experimental data are from LHCb\cite{2}.}
\label{fig9}
\end{center}
\end{figure}

\begin{figure}
\begin{center}
\includegraphics[width=7.6cm]{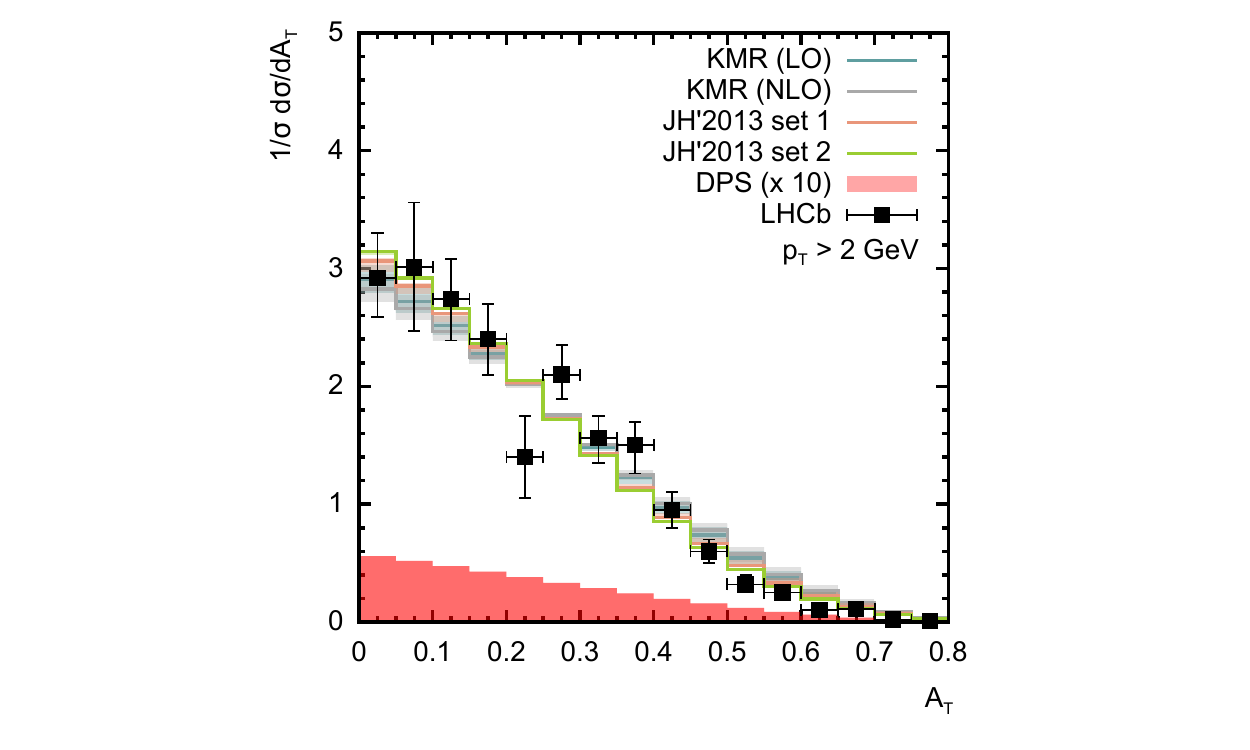}
\includegraphics[width=7.6cm]{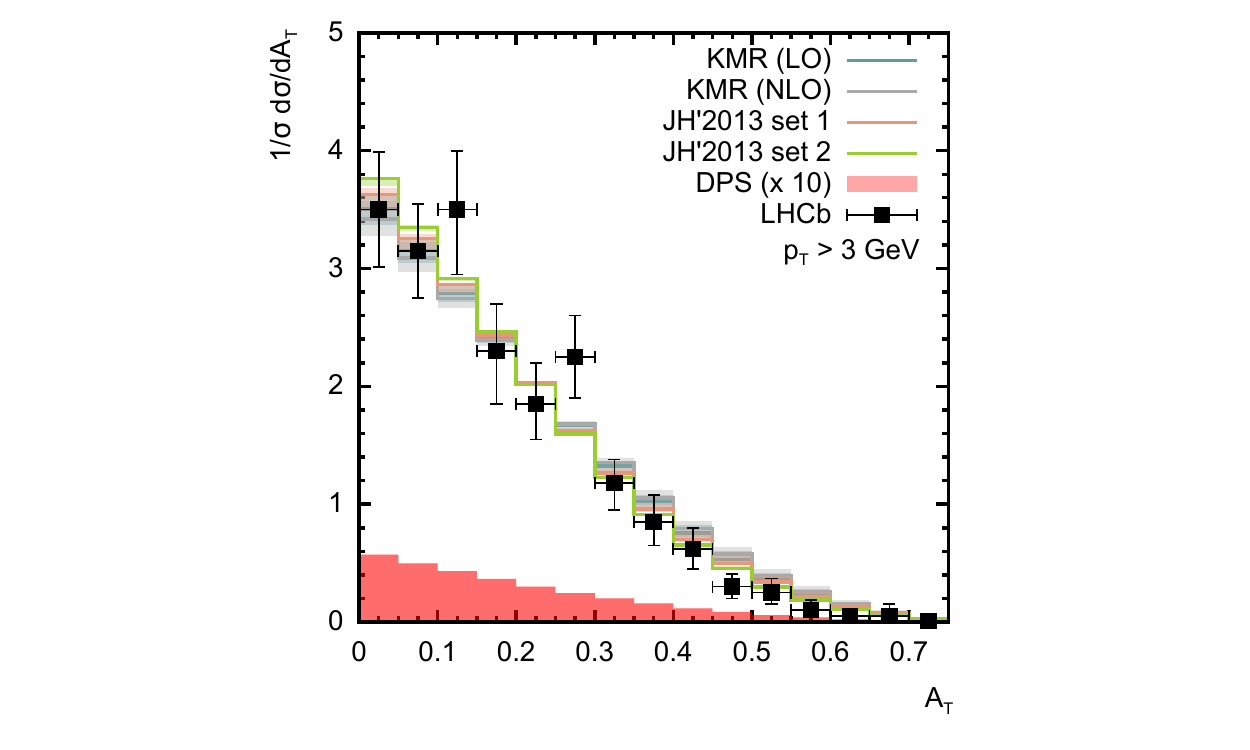}
\includegraphics[width=7.6cm]{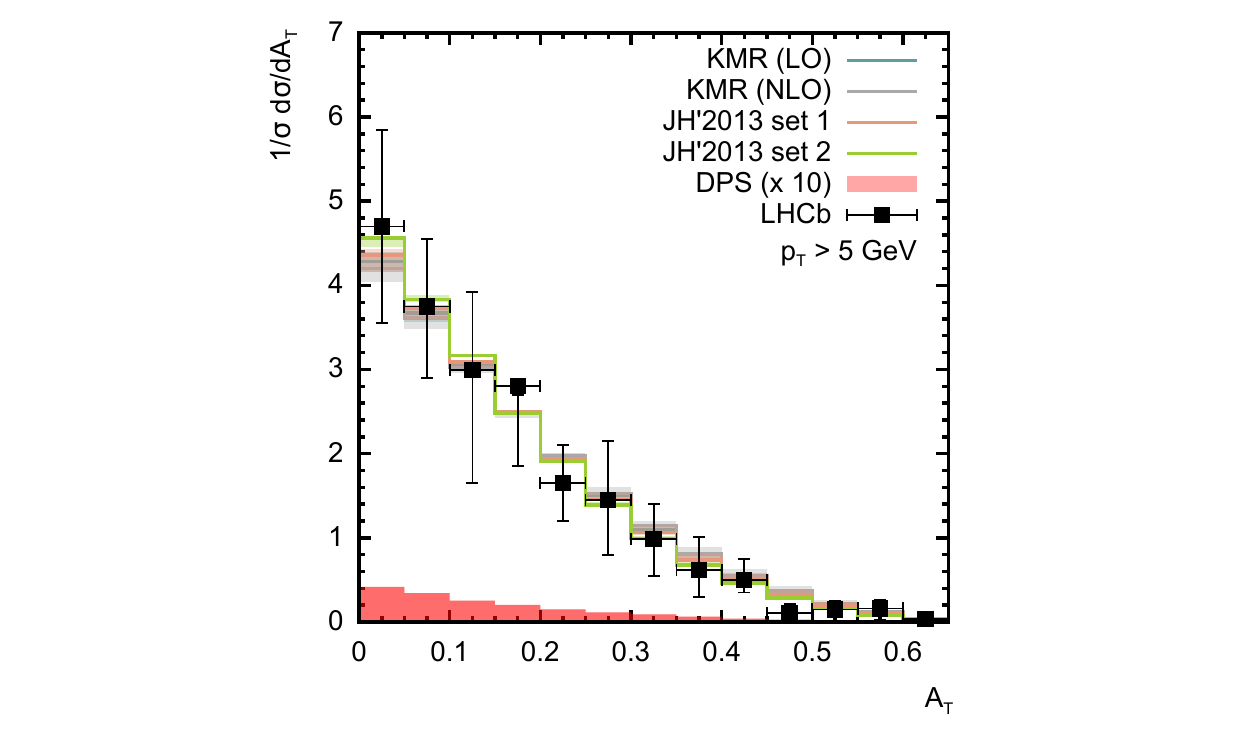}
\includegraphics[width=7.6cm]{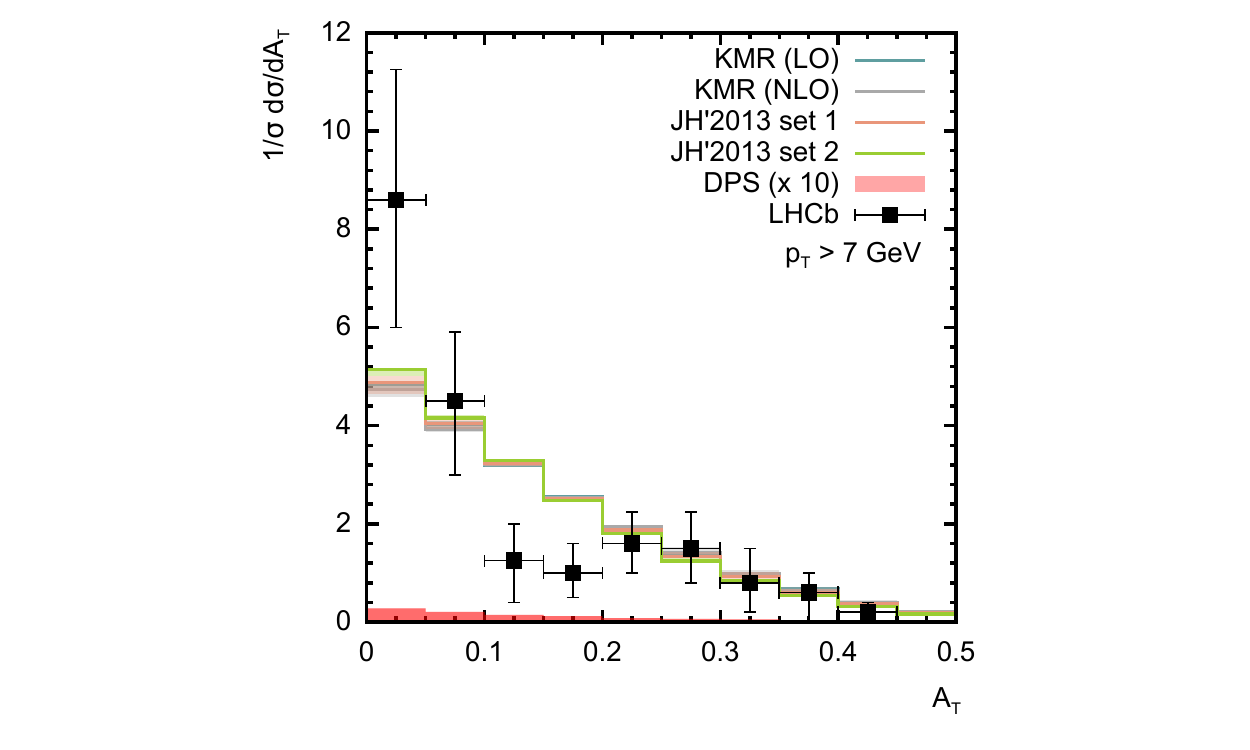}
\caption{The normalized differential cross sections of non-prompt $J/\psi + J/\psi$ 
production at $\sqrt s = 8$~TeV as a function of assymetry
${\cal A}_T$ between the transverse momenta of two $J/\psi$ mesons.
Notation of histograms is the same as in Fig.~1. The experimental data are from LHCb\cite{2}.}
\label{fig10}
\end{center}
\end{figure}

\begin{figure}
\begin{center}
\includegraphics[width=7.6cm]{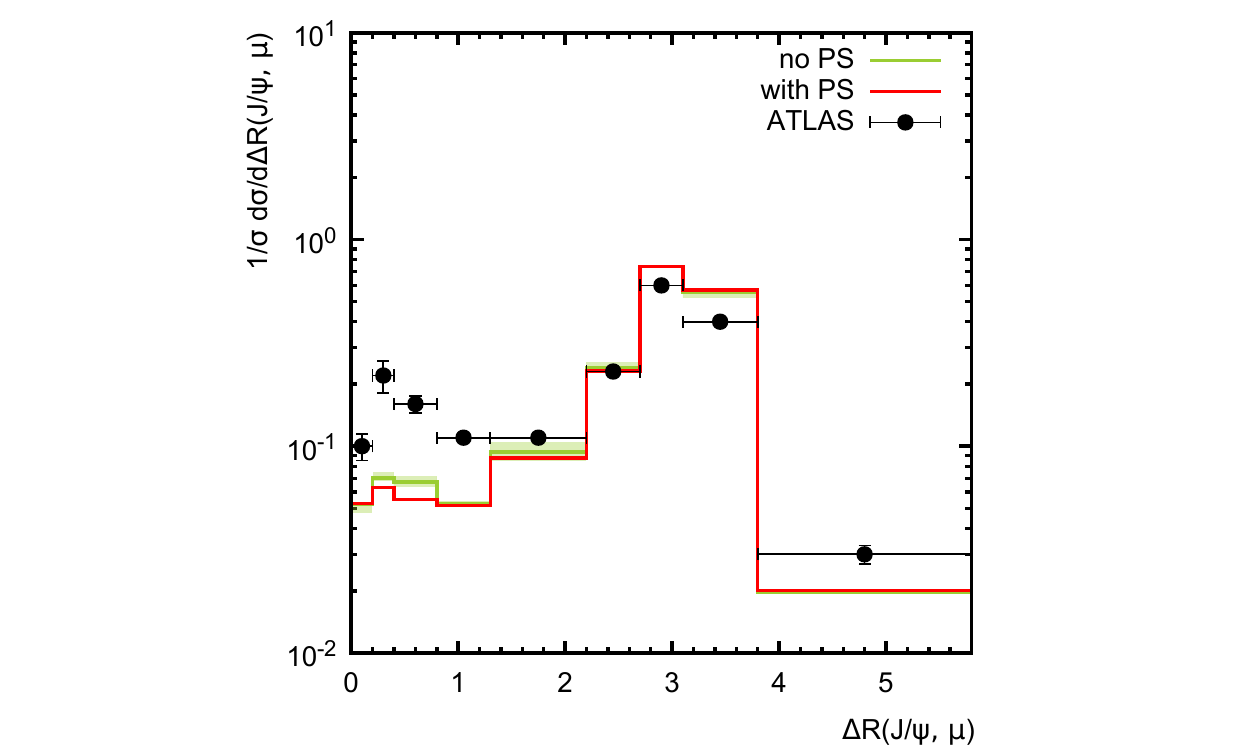}
\includegraphics[width=7.6cm]{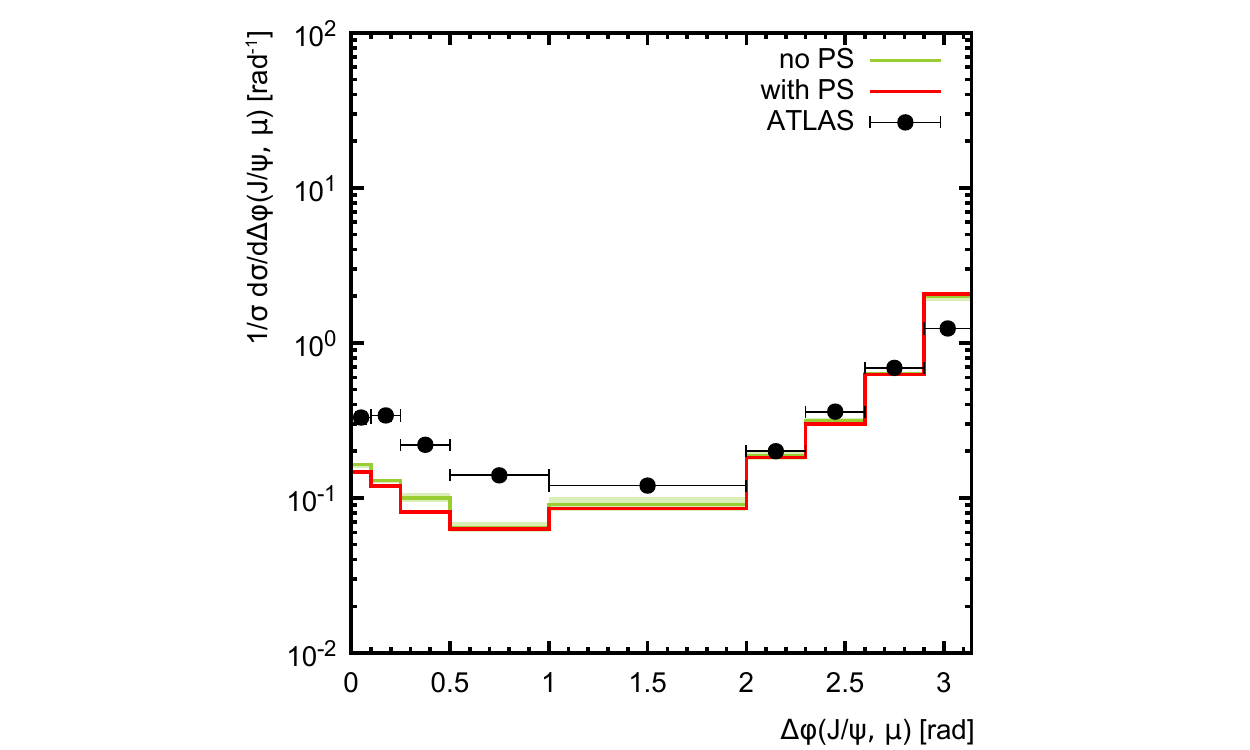}
\includegraphics[width=7.6cm]{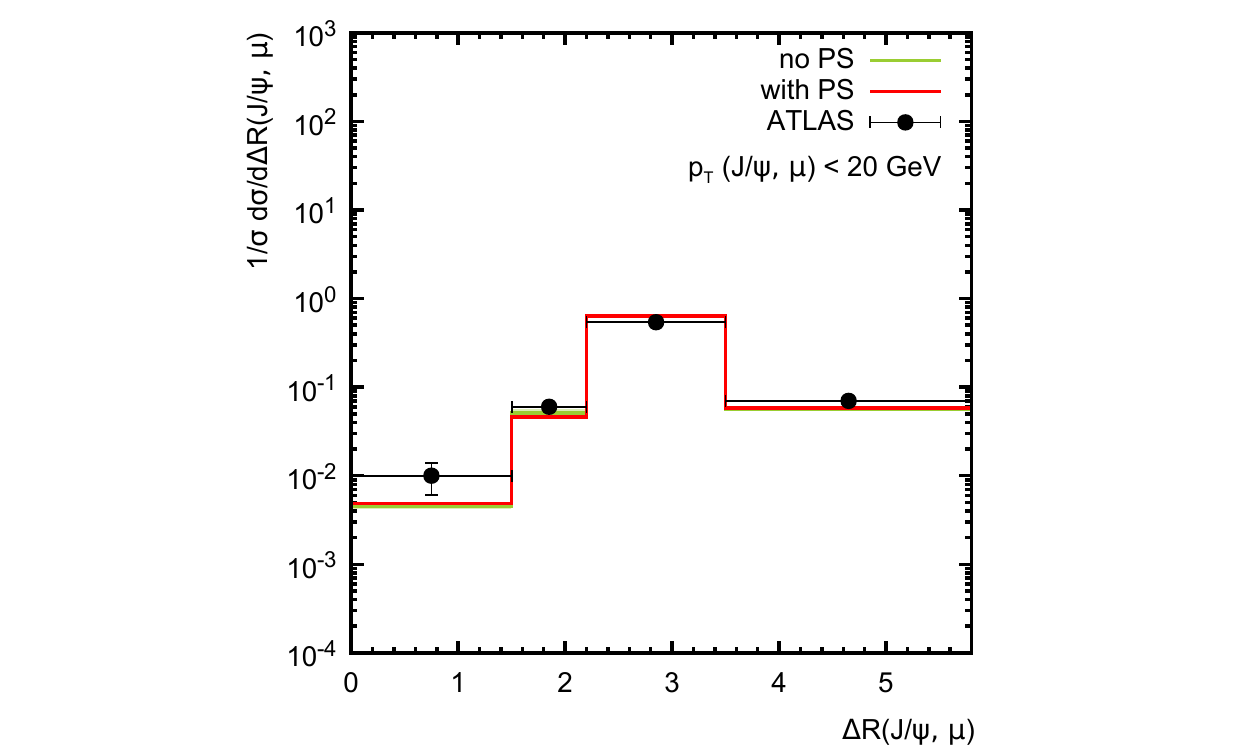}
\includegraphics[width=7.6cm]{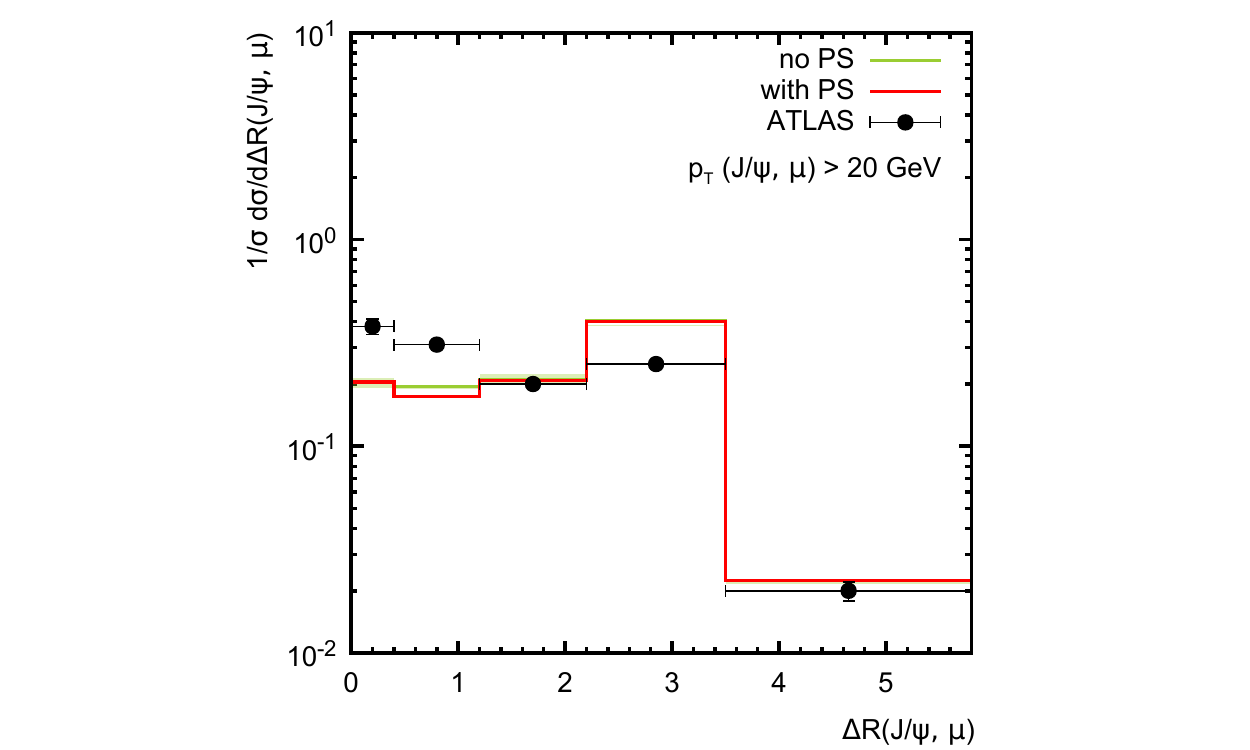}
\caption{Influence of the parton shower effects on the angular
correlations in the associated non-prompt $J/\psi + \mu$ production
at $\sqrt s = 8$~TeV. The {\it JH'2013 set 2} gluon density is applied.
The experimental data are from ATLAS\cite{1}.}
\label{fig11}
\end{center}
\end{figure}

\begin{figure}
\begin{center}
\includegraphics[width=7.6cm]{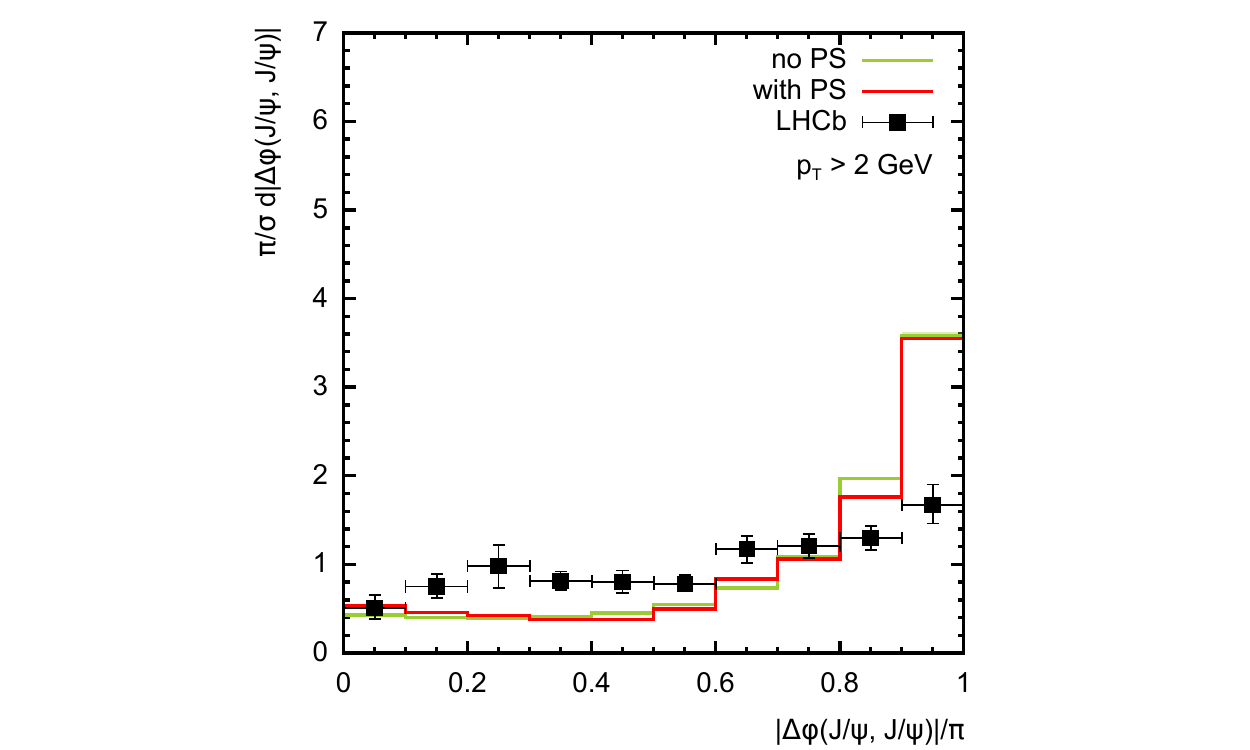}
\includegraphics[width=7.6cm]{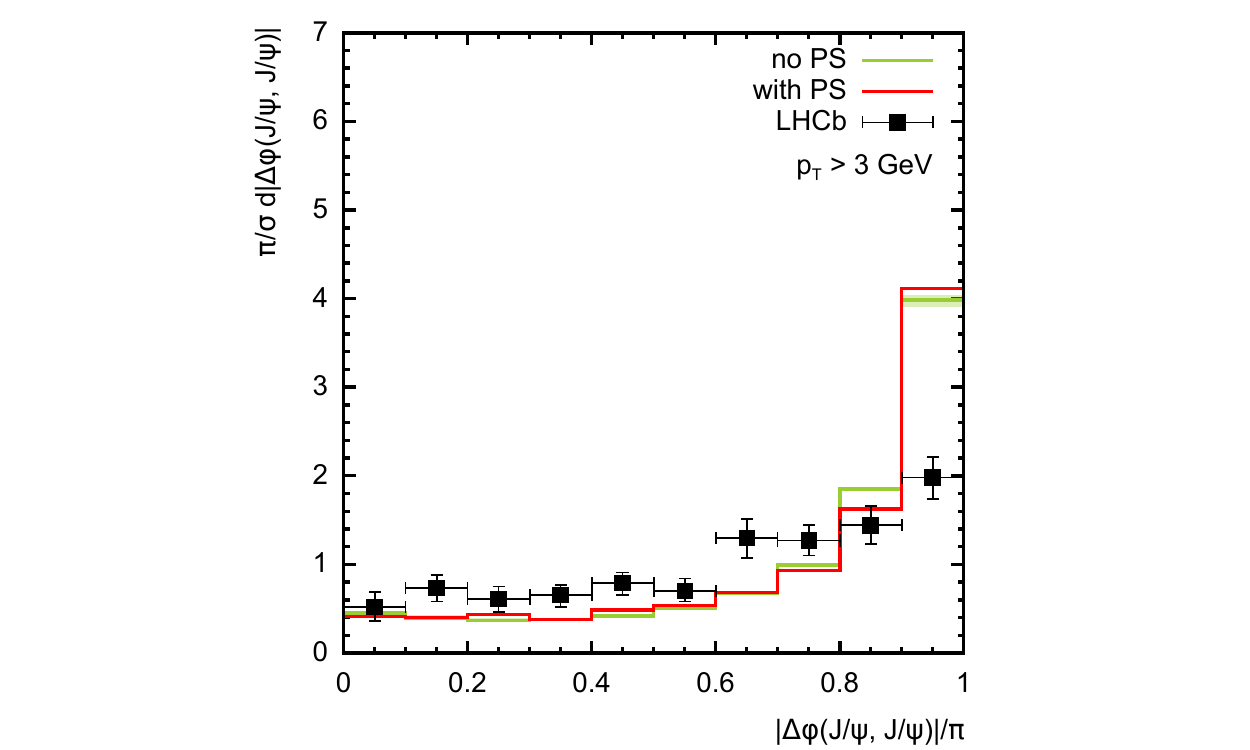}
\includegraphics[width=7.6cm]{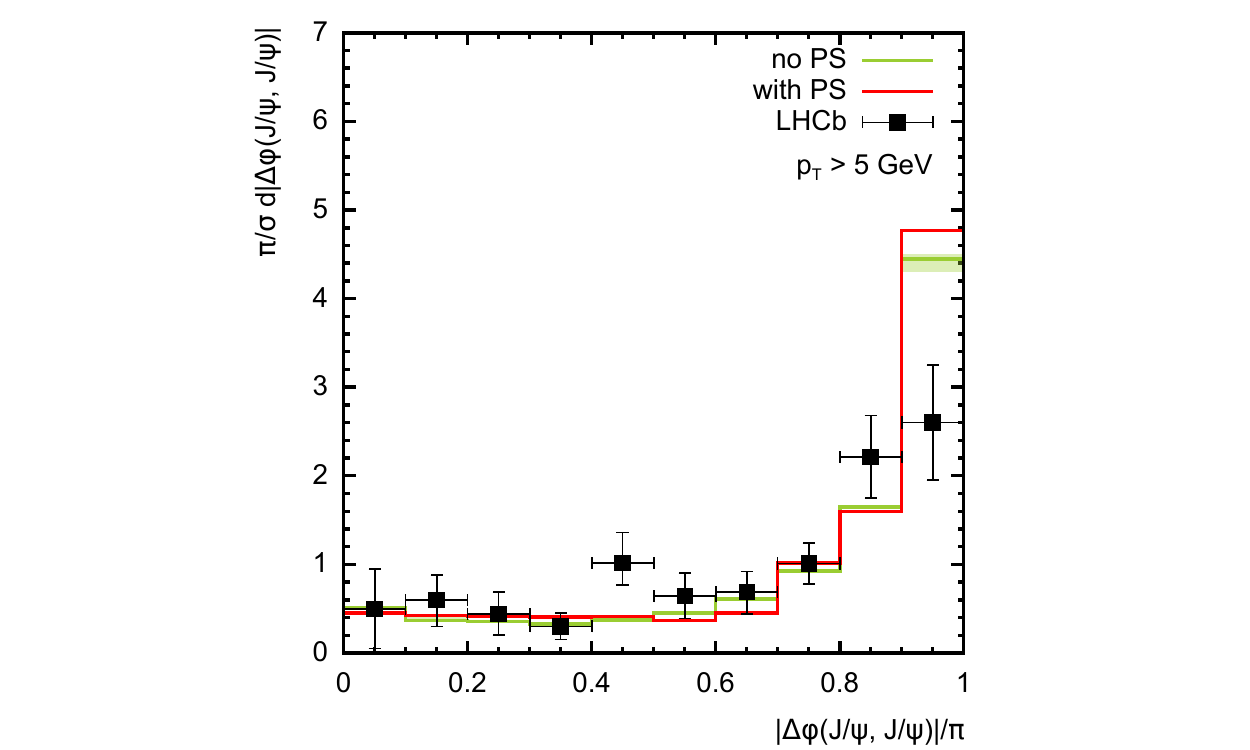}
\includegraphics[width=7.6cm]{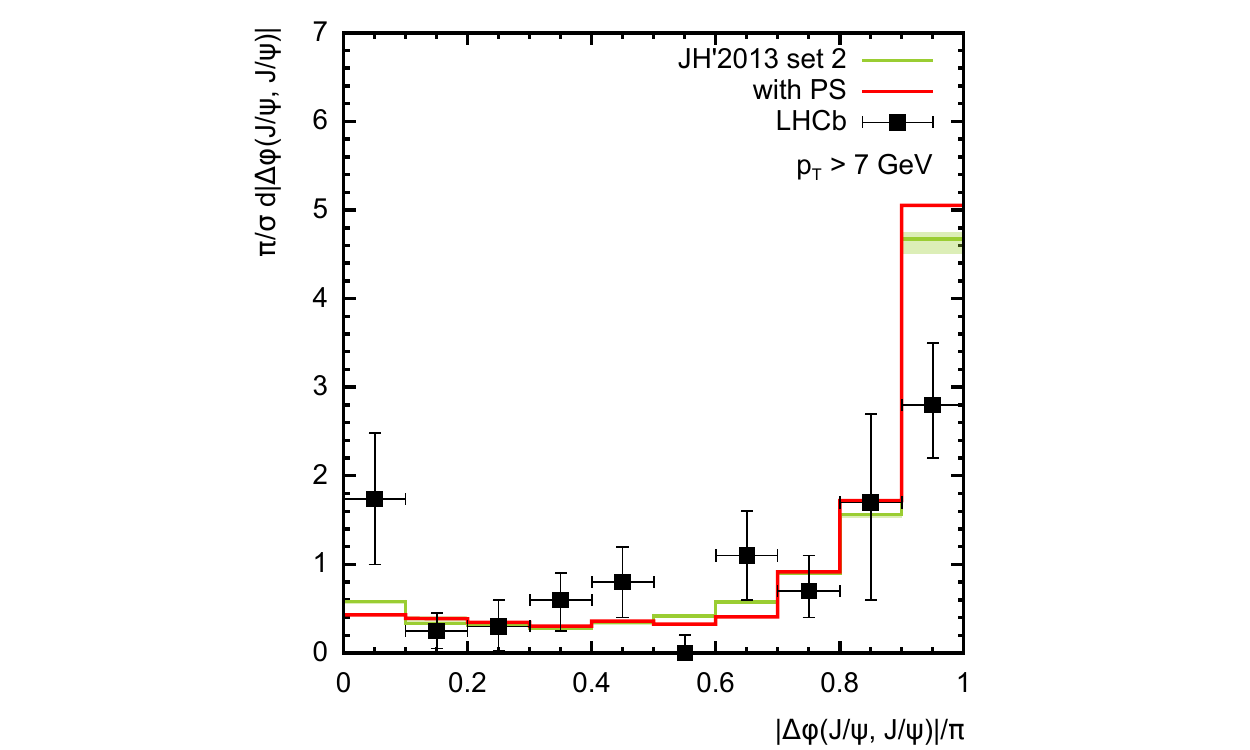}
\caption{Influence of the parton shower effects on the angular
correlations in the non-prompt $J/\psi + J/\psi$ production
at $\sqrt s = 8$~TeV. The {\it JH'2013 set 2} gluon density is applied.
Notation of histograms is the same as in Fig.~11.
The experimental data are from LHCb\cite{2}.}
\label{fig12}
\end{center}
\end{figure}

\end{document}